\documentclass[twocolumn]{aastex631} 
\usepackage{enumitem}
\usepackage{xcolor}
\usepackage{amsmath}
\usepackage{enumitem}
\renewcommand\labelenumi{(\roman{enumi})}
\renewcommand\theenumi\labelenumi
\usepackage{xcolor, soul}
\sethlcolor{yellow}
\usepackage{subfigure}
\usepackage{tablefootnote}
\renewcommand{\theenumi}{\Roman{enumi}}
\usepackage{bm}
\hypersetup{backref=true,pagebackref=true,  hyperindex=true,colorlinks=blue,breaklinks=true,  urlcolor=violet,linkcolor= red, bookmarks=true, 	bookmarksopen=true,  filecolor=cyan, citecolor=blue, linkbordercolor=blue}
\shorttitle{Self-lensing in WD+NS and WD+BH binary systems}
\shortauthors{Sajadian, et. al.}

\begin{document}

\title{Self-Lensing Signals in Binary Systems Containing White Dwarfs with Neutron star or Stellar-mass Black hole Companions}

\author[0000-0002-0167-3595]{Sedighe Sajadian}
\affiliation{Department of Physics, Isfahan University of Technology, Isfahan 84156-83111, Iran, \url{s.sajadian@iut.ac.ir}}

\author{Man Ho Chan}
\affiliation{Department of Science and Environmental Studies, The Education University of Hong Kong, Hong Kong, China}

\begin{abstract}
Light curves from binary systems containing white dwarfs with neutron star or stellar-mass black hole companions (WD+NS and WD+BH) with edge-on orbital planes potentially show self-lensing/eclipsing signals. Here, we evaluate the properties and detectability of these signals in the NASA's Transiting Exoplanet Survey Satellite (TESS), and the Nancy Grace Roman Space Telescope (Roman) observations. WD+NS systems with orbital periods $T\lesssim25~$days mostly have considerable finite-source sizes with the normalized source radii $\rho_{\star}\gtrsim1$. WD+BH systems with $T\gtrsim3$ days have $\rho_{\star}\lesssim1$, and $\rho_{\star}\sim0.01$ for BHs with a few tens solar-mass. Our analytical calculations show the probabilities of occurring self-lensing signals in WD+NS and WD+BH systems are $\sim10^{-3},~10^{-2}$, and maximize for systems with low-mass WDs revolving massive NSs/BHs. We simulate their light curves and generate synthetic data for them by applying the observing protocols of these two satellites. We assume self-lensing signals are detectable if (i) $1\leq T\leq T_{\rm{obs}}$ (where $T_{\rm{obs}}=62~\rm{and}~27.4$ days are the Roman and TESS continuous observing windows), (ii) $\rm{SNR}\ge3,~6$, their signals are (iii) deeper than twice the photometric error, and (iv) covered by at least one datum. Systems with detectable self-lensing signals in the TESS and Roman observations on average have small inclination angles $i\lesssim0.2^{\circ}$, with the orbital periods $\sim6,~19~$days, and their signals last $\sim[6,~30]~\rm{minutes}$. The TESS and Roman efficiencies for detecting these signals are $\sim2-6\times10^{-4}$ and $\sim2-12\times10^{-10}$. Although detecting these self-lensing signals by Roman is impossible, the TESS telescope potentially manifests at least one self-lensing signal due to these binary systems, if $8\%,~\rm{and}~3\%$ of WDs have NS and BH companions.         
\end{abstract}


\section{Introduction}\label{sec1}
Binary systems composed of compact objects (including white dwarfs, neutron stars, and black holes) are potential sources of novae explosions, Gamma-ray bursts, and gravitational waves, earning them the title of multi-messenger targets. When two components of these binary systems merge, they emit gravitational waves, which are ripples in space-time, predicted by Einstein's General Theory of Relativity \citep{1916Einstein,1918Einstein}. In September 2015, the first direct detection of gravitational waves was recorded by the Laser Interferometer Gravitational-Wave Observatory (LIGO, \citet{2015LIGOcollaboration}) detectors \citep{2016PhRvFirstGW}. Binary neutron star systems are the most common sources of gravitational waves. At the time of their merging, they produce kilo-novae, short Gamma-ray bursts, and electromagnetic signals as well. For instance, the recorded gravitational wave GW170817 by the LIGO and Virgo \citep{2015Virgocollaboration} detectors simultaneously in 2017 originated from the merger of two neutron stars \citep{PhysRevLettGW170817}.

In addition, the merger of compact objects can lead to a nova explosion. For example, a Type-Ia supernova occurs when two carbon-oxygen white dwarfs merge \citep[see, e.g., ][]{2010NaturTypeIaSup,2012AAToonen}. More massive compact objects at the merging time convert to more luminous novae. When compact objects in binary systems merge and during a supernova event, an explosion takes place that emits gamma rays and a longer-lasting afterglow, known as a Gamma-ray burst (GRB). These energetic cosmic events can be classified into two types according to their durations: long-duration (lasting longer than approximately 2 seconds) and short-duration (lasting less than 2 seconds) \citep[see, e.g.,][]{GRBreview2004,2018bookzhang}. Short-duration GRBs primarily originate from the merger of two neutron stars in close binary systems, while long-duration GRBs are typically emitted due to the death of a massive star, although some of them may occur during the merger of compact objects \citep{2022NaturGRBs,2025ApJWang,2023ApJGotttlieb}.

Detecting and studying binary systems composed of compact objects are crucial for several reasons: testing Einstein's General Theory of Relativity,  studying dense matter physics, probing potential sources of GRBs and gravitational waves, etc. \citep{2002ApJBCOs,2006ApJbelczynski}. White dwarfs (WDs) in binary systems with other compact objects are common, and some of them are detectable photometrically depending on their apparent brightness. The estimated number of WD-neutron stars (WD+NS) or WD- stellar-mass black hole (WD+BH) binary systems in our galaxy is in the order of a few million \citep[see, e.g., ][]{2006PhRvDNSWDN,2024MNRASkorol}, while the number of Galactic double WD (DWD) systems is as high as 100-1000 million \citep{2017AAToonen}. Spectroscopic survey observations by the Sloan Digital Sky Survey (SDSS) and the Gaia telescope have revealed binary systems from compact objects including at least one WD, i.e., DWD, WD+NS, WD+BH, \citep[see, e.g.,][]{2009ApJWDNS,2020ApJSDSSDWD98,2025arXivSDSSDWD,2025chawla}. For these systems, follow-up photometric observations from such systems will manifest periodic signals or trends in their light curves with various origins, such as self-lensing \citep{2002ApJAgolSelfL}, eclipsing, Doppler-boosting \citep{2003ApJLoebDoppler}, ellipsoidal variation \citep{1993ApJMorrisEllipsoidal}, heating and irradiation \citep{1992MNRASDaveyirradiatin,2002ASPCBeerIrradiation,2004AAclaretheating}, etc. \citep[see, e.g.,][]{2025AJsajadianasadi}. Detecting and characterizing these periodic trends and signals helps to resolve degeneracy and measure their physical parameters \citep{2024ApJsorebella}. However, studying and modeling these variations before searching for them will give us insights about their amplitudes, durations, and detectability.

Self-lensing refers to a temporary lensing effect in binary systems that are positioned edge-on as seen by the observer. Measuring these periodic signals in the light curves of binary systems is possible in dense and accurate photometric data, such as by the NASA' Transiting Exopanet Survey Satellite (TESS) telescope \citep{2014SPIETESS,2015JATISRicker,2016SPIETTESS}. In this study, we numerically investigate the properties and detection of self-lensing signals in WD+NS and WD+BH binary systems. By assuming edge-on orbital planes for these systems, we simulate their light curves. We then apply two observing strategies, and determine whether their self-lensing signals can be identified in real data.

The outline of this paper is as follows. In Section \ref{sec2}, we will explain our formalism for simulating self-lensing signals in WD+NS and WD+BH binary systems. We analytically evaluate some properties of self-lensing signals in such systems and their detectability. We then perform Monte Carlo simulations from these systems, create their light curves, and generate synthetic data points (taken by TESS and Roman) for them, as will be explained by details in Section \ref{sec3}. In Subsection \ref{sub33}, we study the detection rate of their self-lensing signals. In Section \ref{snew}, we will estimate the number of detectable self-lensing signals from these systems in the TESS and Roman observations. In Section \ref{sec5}, we will explain the results and conclusions.

\section{Self-Lensing in WD+NS and WD+BH Binary Systems: Analytical Approach}\label{sec2}
The subject of periodic lensing signals in binary systems was first studied by \citet{1973AAMaeder}, where he concluded that self-lensing in binary systems with two compact objects is $10$-$25$ times larger than in binary systems with one compact object and one main-sequence star. The lensing optical depth and occultation effects (or finite-lensing effects) in self-lensing signals were theoretically and numerically investigated by \citet{1995ApJGouldSelfL,2001MNRASMarsh,2002ApJAgolSelfL,2016ApJHan,2025AJSajadian}. Also, \citet{2002AABeskin} investigated the detection of self-lensing in binary systems with compact objects through SDSS observations. Based on Kepler photometric data, self-lensing features in five binary systems including WDs and main-sequence stars (WD+MS) were found \citep{2014ScienceKruse,2019ApJMasuda,2018AJSelfLKawahara}. Recently, for one of these targets its self-lensing signal was re-detected through the TESS observations \citep{2024ApJsorebella}. The potential detection of self-lensing signals in observations by TESS, Roman, and the Vera C. Rubin Observatory has been studied by \citet{2017MNRASKorol,2019ApJMasuda,2021MNRASWiktorowicz,2025Wikrorowicz,2025AJSelfDWD}. As a result, based on the realistic simulations, \citet{2024AJSajadianAfshordi} predicted that the TESS telescope would detect self-lensing signals in approximately $15$ WD+MS and $6$ NS+MS systems. Here, we aim to extend this research and specifically focus on self-lensing signals in WD+NS and WD+BH binary systems. In the following subsection, we review our formalism for generating self-lensing signals in such binary systems.  
\begin{figure*}
\centering
\includegraphics[width=0.32\textwidth]{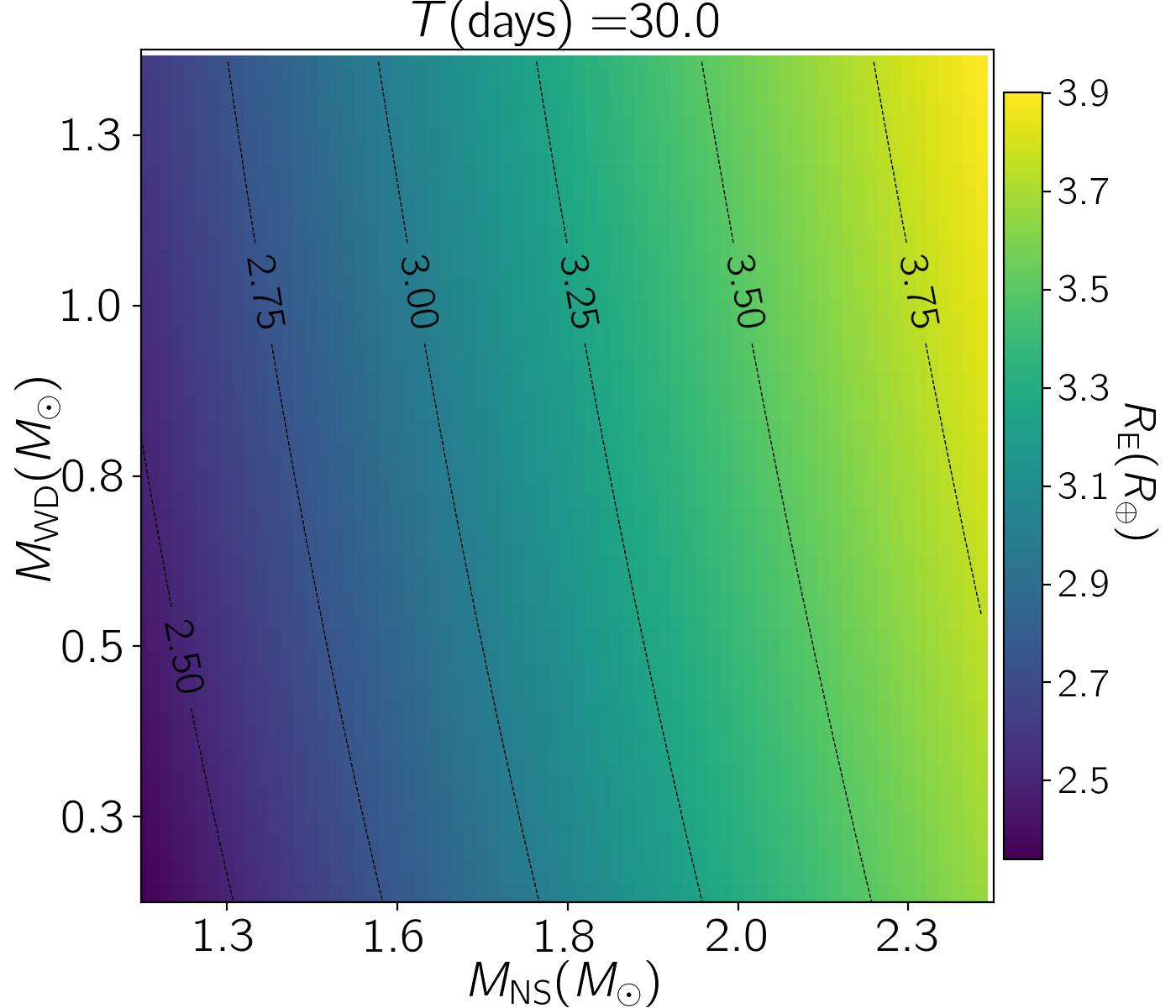}
\includegraphics[width=0.32\textwidth]{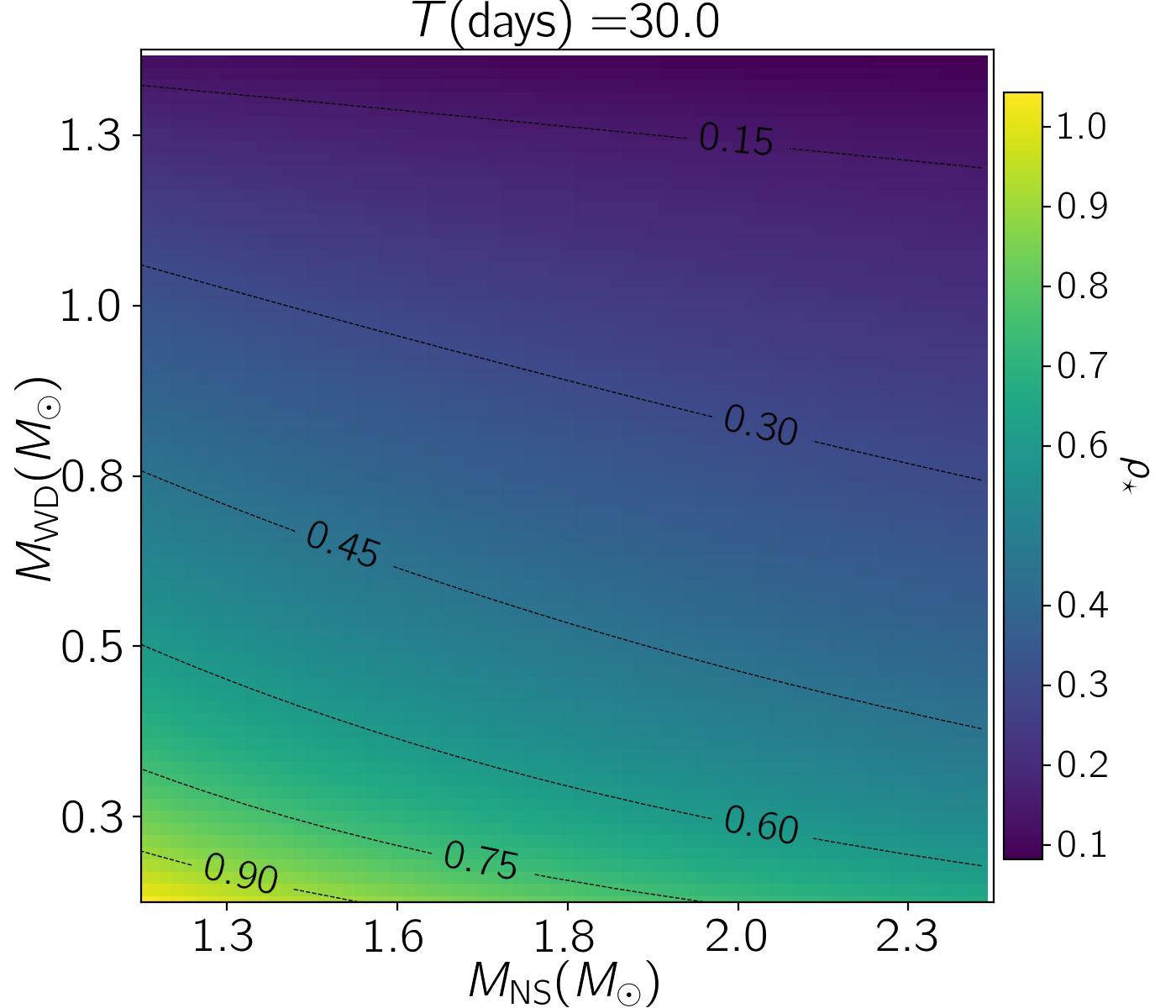}
\includegraphics[width=0.33\textwidth]{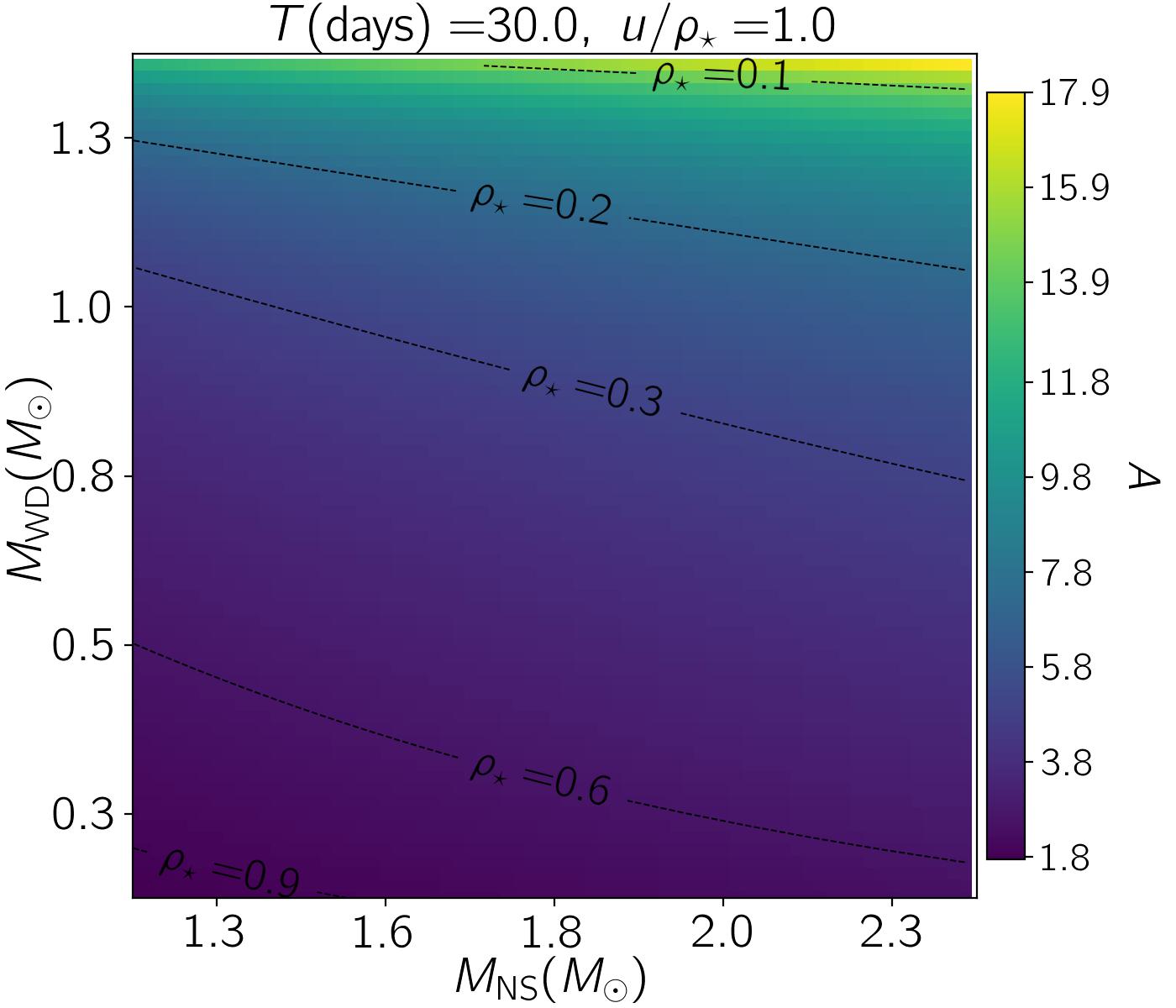}
\includegraphics[width=0.32\textwidth]{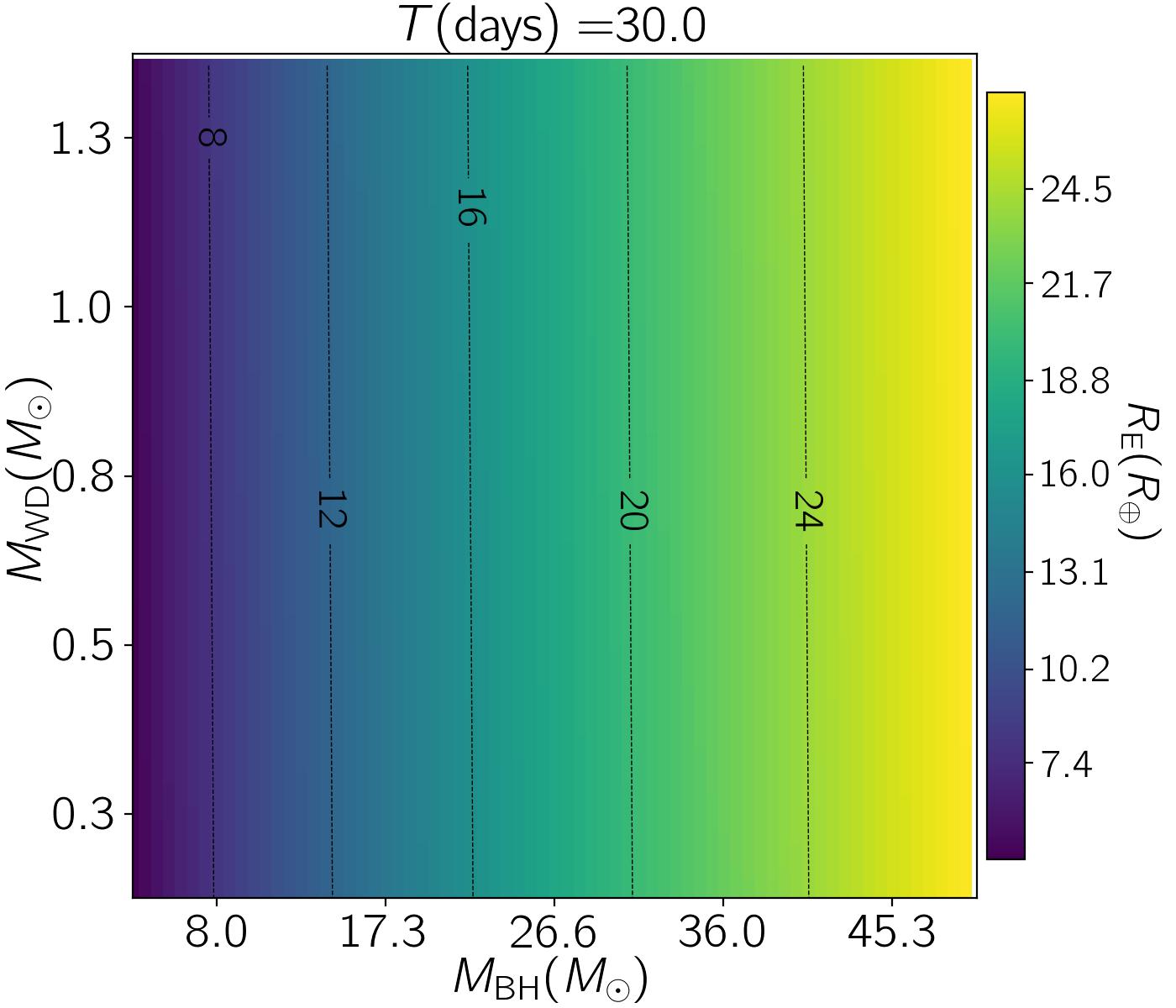}
\includegraphics[width=0.32\textwidth]{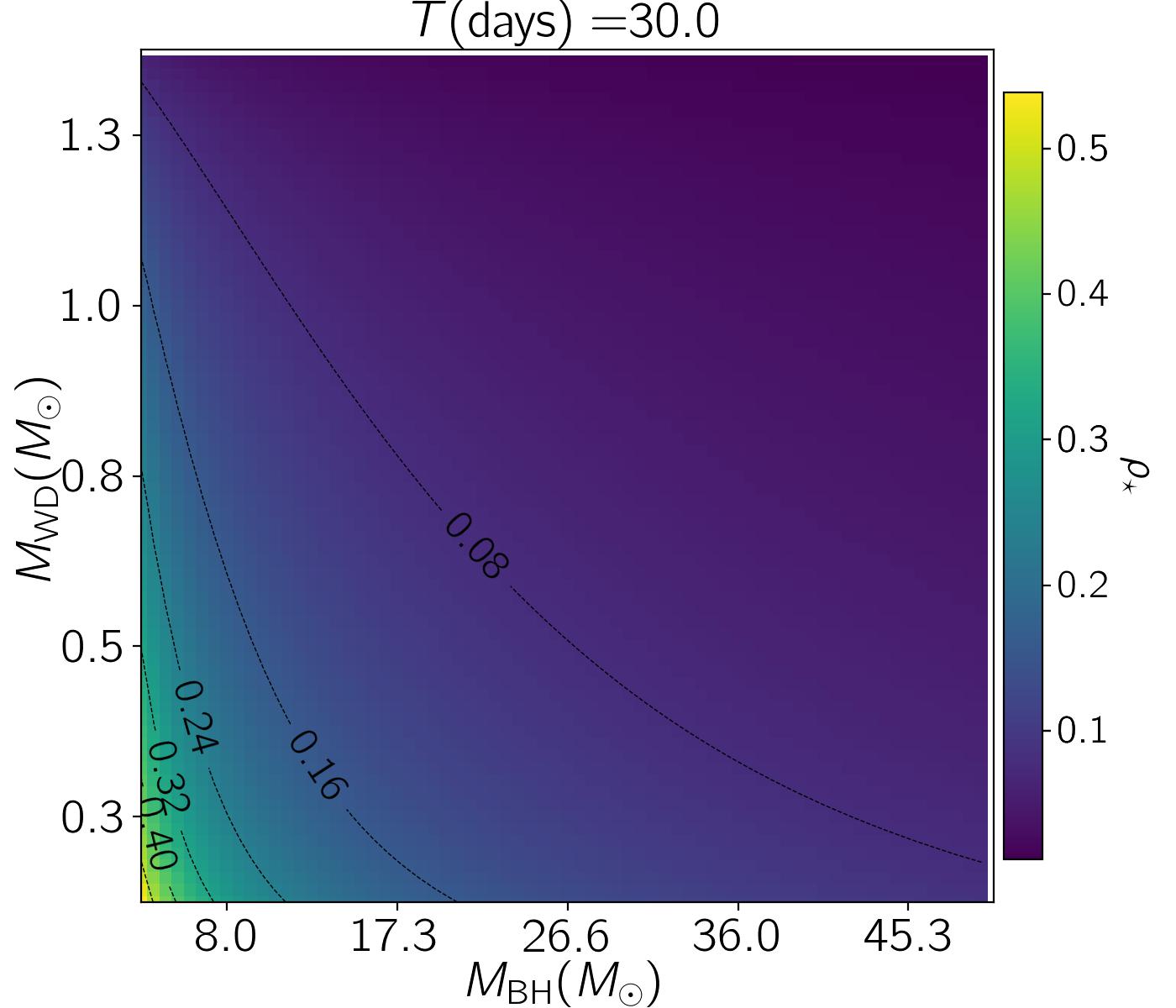}
\includegraphics[width=0.33\textwidth]{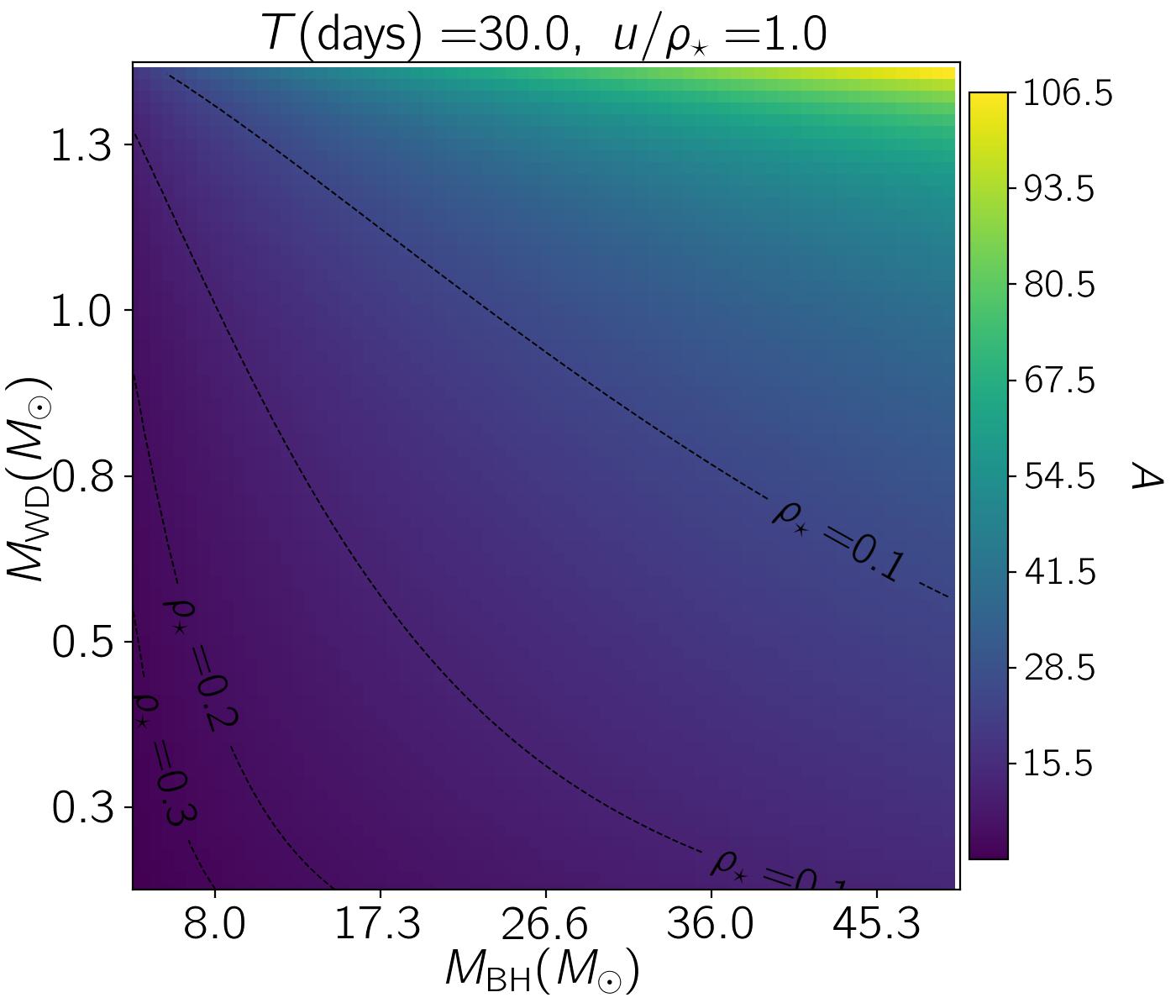}
\caption{Maps of $R_{\rm E}$ (left panels), $\rho_{\star}$ (middle panels), and $A$ the magnification factor for binary systems WD+NS (top row) and WD+BH (bottom row) over the 2D space made from $M_{\rm{WD}}$ (vertical axes), and mass of NSs and BHs (horizontal axes). Here, we assume WDs are source stars, while NSs or BHs are the lens objects. Similar maps to these maps but for other values of orbital periods are also available in \href{https://iutbox.iut.ac.ir/index.php/s/QzKPgdKTASLromx}{link1}, \href{https://iutbox.iut.ac.ir/index.php/s/GHwrgccjNQ7rLB6}{link2}, \href{https://iutbox.iut.ac.ir/index.php/s/rdwbwjXGd4pqbNP}{link3}, \href{https://iutbox.iut.ac.ir/index.php/s/MzGdgdeXRaKTtWF}{link4},
\href{https://iutbox.iut.ac.ir/index.php/s/YKfE7BY2RJQpa3T}{link5}, and \href{https://iutbox.iut.ac.ir/index.php/s/RqYt2YHDreygBQw}{link6}, respectively.}
\label{fig1}
\end{figure*}

\subsection{Formalism}\label{sub21}
To formulate WD+NS and WD+BH binary systems we introduce the following parameters: the mass of the WD as the source star $M_{\star}$, the mass of the NS or BH as the lens object $M_{\rm l}$, their orbital period $T$, the orbital eccentricity $\epsilon$, the inclination angle of their orbital planes with respect to the line of sight toward the observer $i$, and the projection angle between the orbital semi-major axis and the sky plane $\theta$. The semi-major axis ($a$) is determined using Kepler's third law.  

We assume that there are no external forces acting on the binary system, and so its components revolve around their common center of mass. In order to determine the location of the binary components versus time we first calculate the mean anomaly, denoted as $\phi=\omega(t-t_{\rm p})$, and then the eccentric anomaly $\xi$, where their relationship is given by the Kepler Equation, i.e., $\phi=\xi-\epsilon\sin\xi$. Here, $\omega= 2\pi/T$ represents the angular velocity, and $t_{\rm p}$ is the time of periapsis. We solve the Kepler equation numerically using Bessel polynomials \citep{1998AADominik,2017ApJSajadianHundertmark}. Next, we need to project the orbital plane onto the sky. That is achieved by applying two rotations: one around the axis normal to the orbital plane by $\theta$, and then another around the intersection axis of the orbital plane and the sky plane by $i$.

At any given time, we calculate the phase angle ($\Phi$) between two lines of sight from the white dwarf toward the observer and the lens object (NS or BH). When the WD passes behind the lens object ($\Phi\in[0,~90^{\circ}]$), its light can either be magnified or blocked by the lens object. However, the occultation amount is small for NS or BH lens objects because of their small sizes. We calculate the lensing effect using a combination of the Python package \texttt{VBMicrolensing} \citep{2010Bozzavbm}, and the inverse-ray shooting method \citep{1992bookSchneider} and utilizing the lens equation as:  
\begin{eqnarray}
\tan\beta=\tan\nu -\frac{D_{\rm{ls}}}{D_{\rm s}}\Big[\tan \nu+\tan(\alpha-\nu)\Big], 
\end{eqnarray}
where, $D_{\rm ls}=D_{\rm s}-D_{\rm l}$, $D_{\rm l}$ and $D_{\rm s}$ are the lens and source distances from the observer, respectively. $\beta$ and $\nu$ are the angles of lines of sight toward the source star and its images with respect to the lens's line of sight, respectively. Also, $\alpha=4GM_{\rm l}/c^{2}b$ is the bending angle of light due to the lensing effect, where $c$ is the light speed, $G$ is the gravitational constant, and $b=D_{\rm l}\tan\nu$. When $\Phi>\pi/2$ and the WD is passing in front of its NS companion, the light of NS is magnified or blocked although its footprint in the overall flux is barely recognizable because of faintness of NSs.

In the lensing formalism, there is a lensing length scale which is called the Einstein radius, which is the radius of the images' ring at the time of the complete alignment of lens, source and the observer, as given by: 
\begin{eqnarray}
R_{\rm E}= \sqrt{\frac{4GM_{\rm l}}{c^{2}}~\frac{D_{\rm ls}D_{\rm l}}{D_{\rm s}}}\sim \sqrt{\frac{4 G M_{\rm l}}{c^{2}}~a}. 
\end{eqnarray}
We note that in binary systems with circular orbits $D_{\rm{l}}\sim D_{\rm s}\gg D_{\rm{ls}}$, and $D_{\rm ls}\sim a$. Self-lensing signals in binary systems suffer from finite-source effect \citep{1994ApJGould,1994ApJWitt}. This effect is characterized by $\rho_{\star}=R_{\star}D_{\rm l}/R_{\rm E} D_{\rm s}$ which is the source (WD) radius projected in the lens plane and normalized to the Einstein radius.

In the lensing formalism, the finite-lens effect is characterized by the lens radius normalized to the Einstein radius, i.e., $\rho_{\rm l}=R_{\rm l}/R_{\rm E}$. Against self-lensing signals in DWDs systems that are affected and reduced by finite-lens effect \citep{2025AJSajadian}, self-lensing signals in WD+NS or WD+BH systems do not have considerable finite-lens effects. In the lensing of a WD by its NS companion, $\rho_{\rm l}\lesssim0.01$, while in the lensing of NS by its WD companion this parameter is $\rho_{\rm l}\gtrsim1$. Hence, if the WD crosses in front of its NS companion, it can block its light completely, although this eclipsing effect is rarely recognizable in the overall flux. 

In the two next subsections, based on the introduced formalism we analytically study the characterization and detection of self-lensing signals in WD+NS and WD+BH binary systems.

\begin{figure*}
\centering
\includegraphics[width=0.33\textwidth]{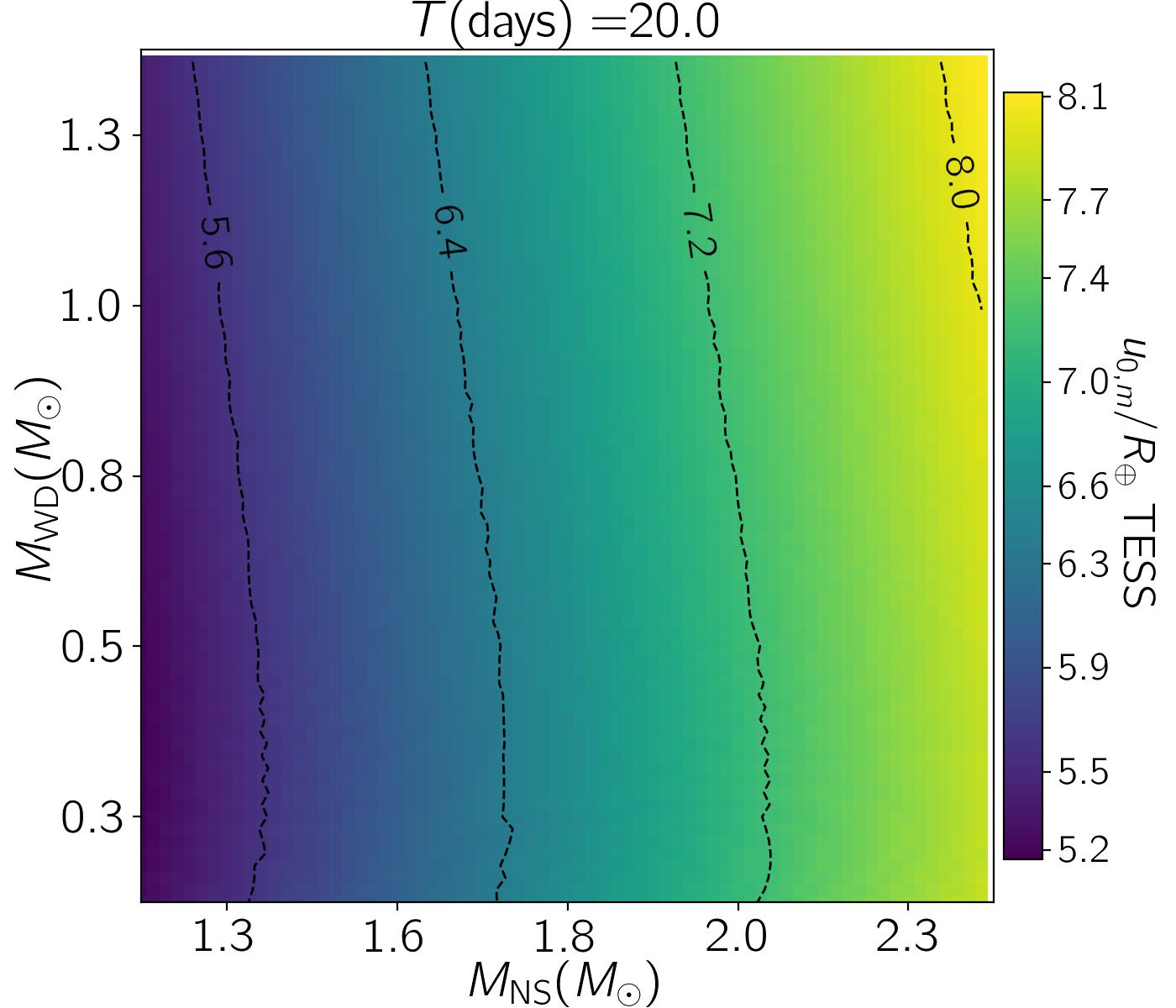}
\includegraphics[width=0.32\textwidth]{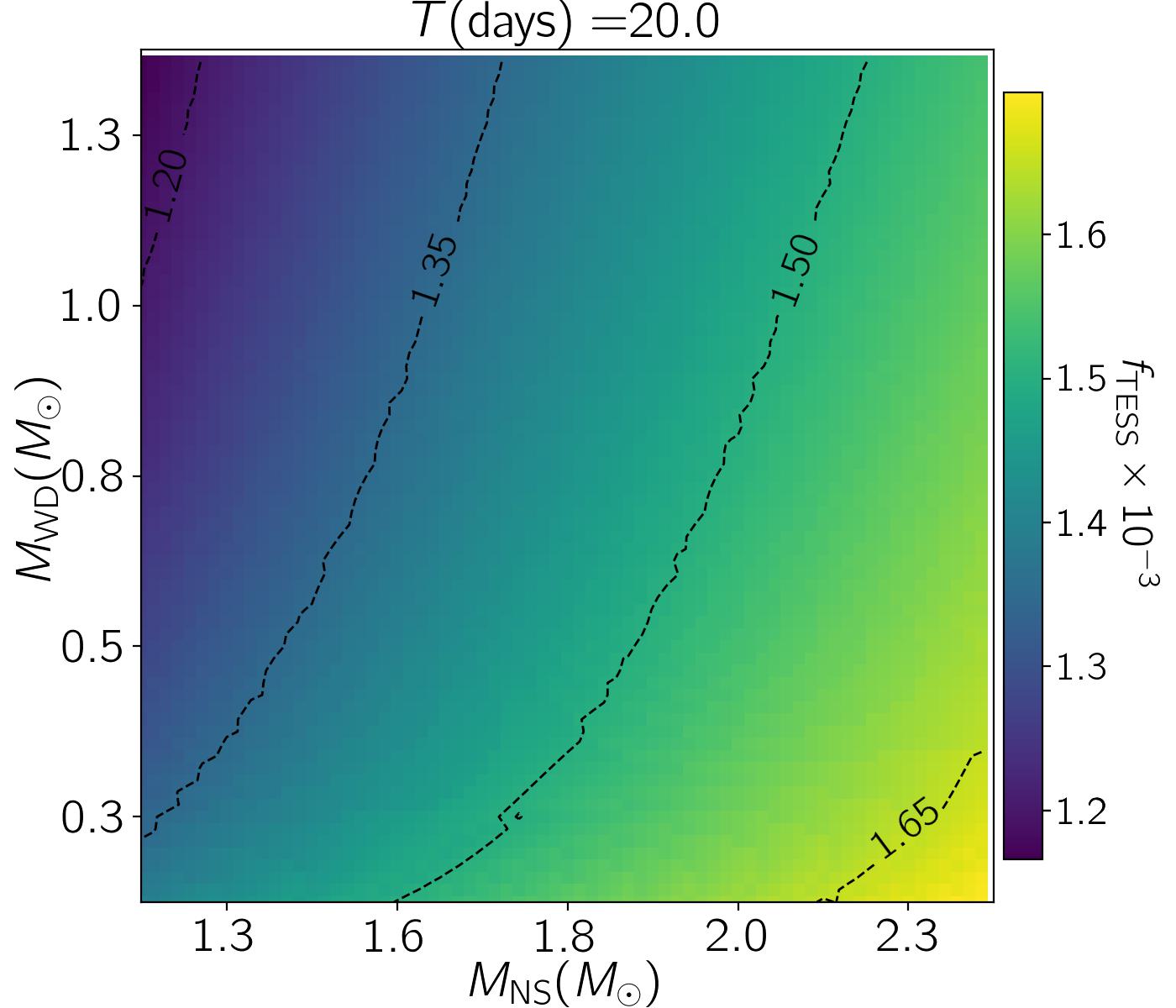}
\includegraphics[width=0.32\textwidth]{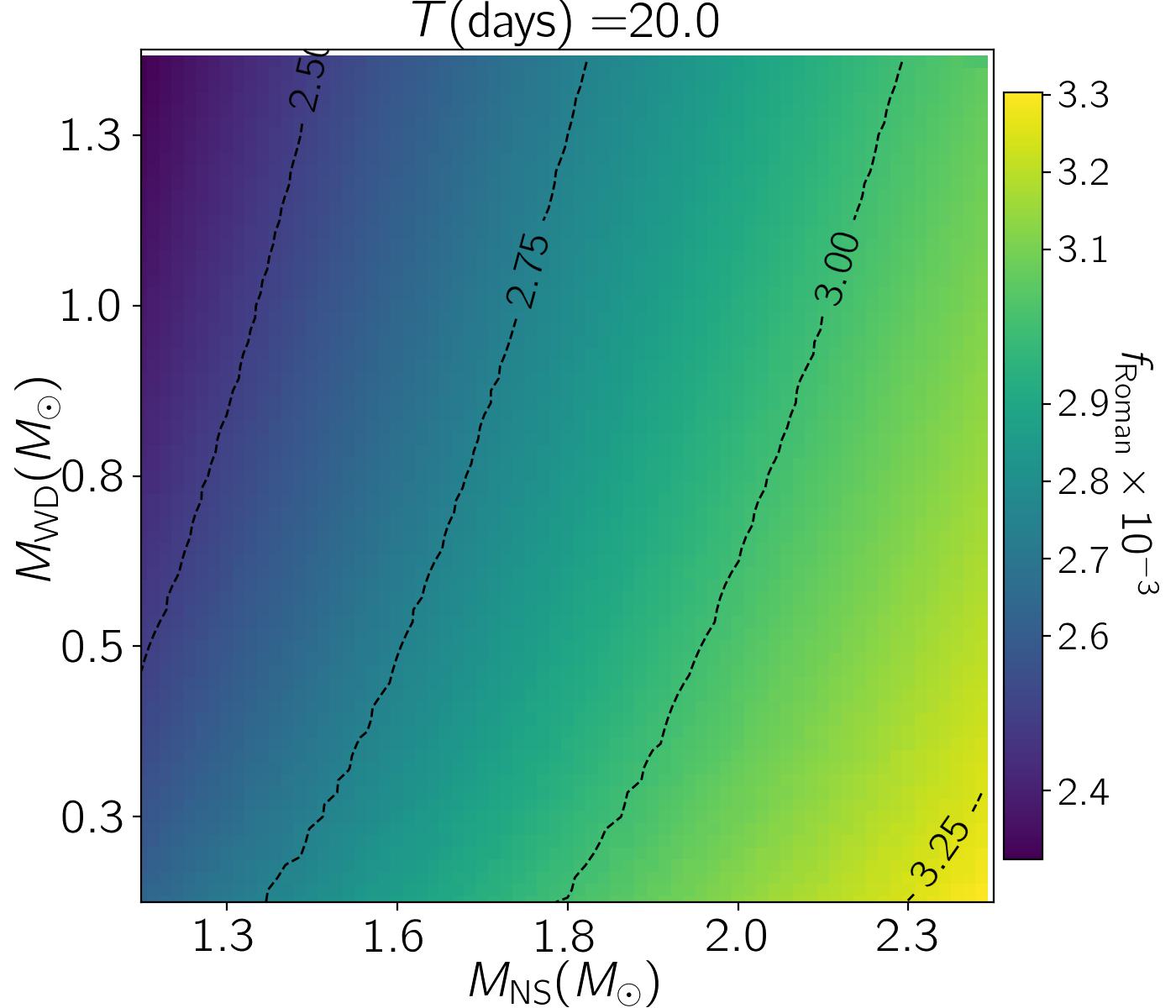}
\includegraphics[width=0.33\textwidth]{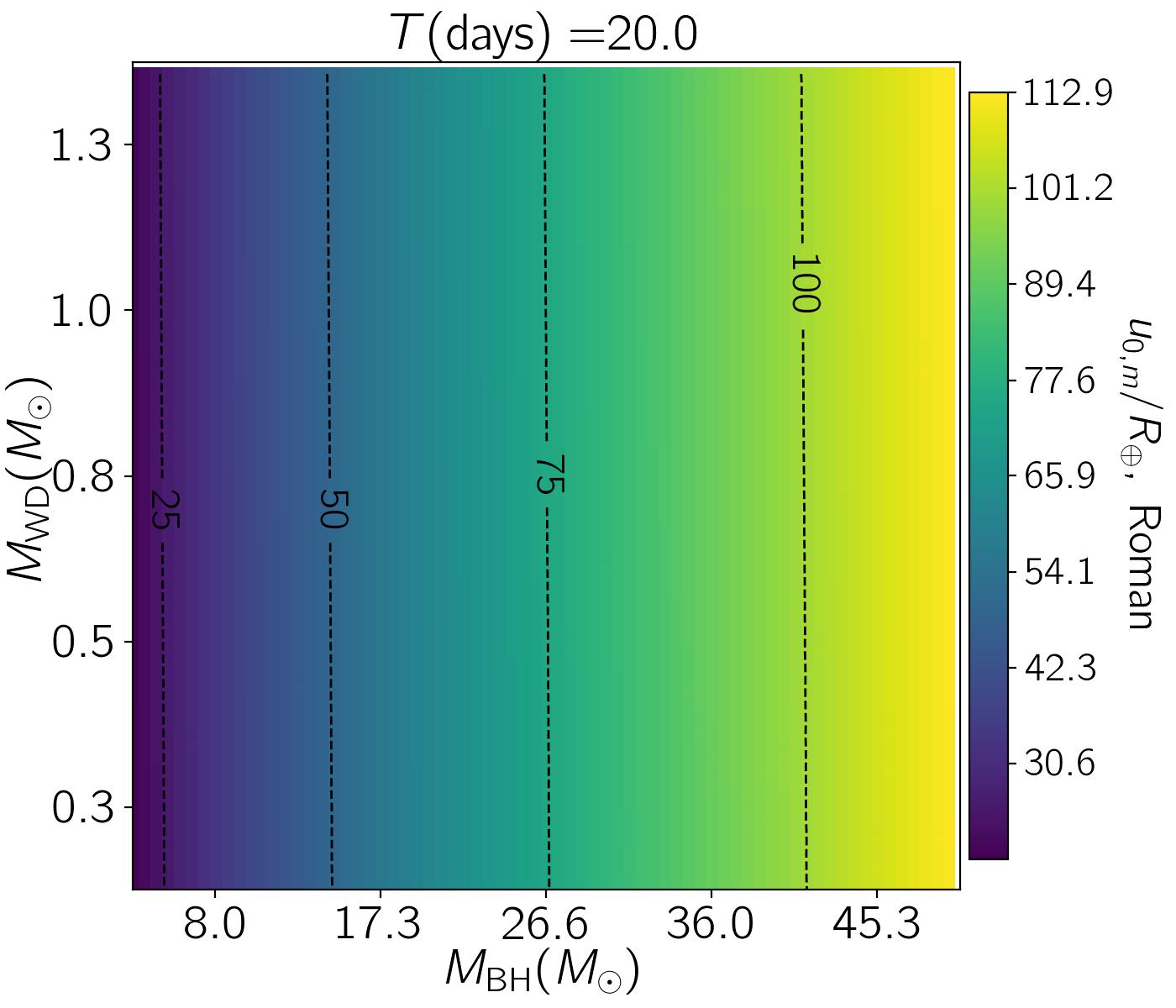}
\includegraphics[width=0.32\textwidth]{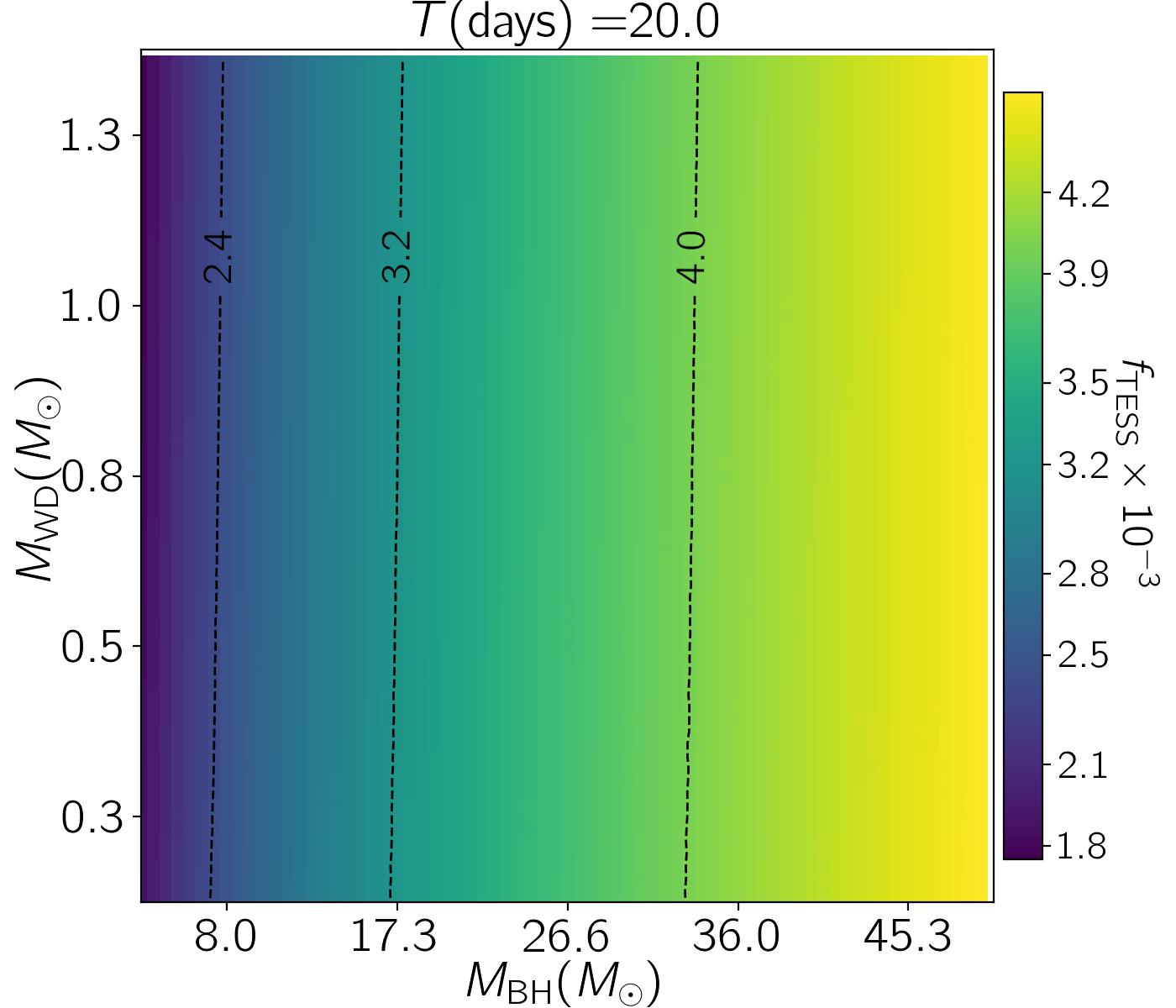}
\includegraphics[width=0.32\textwidth]{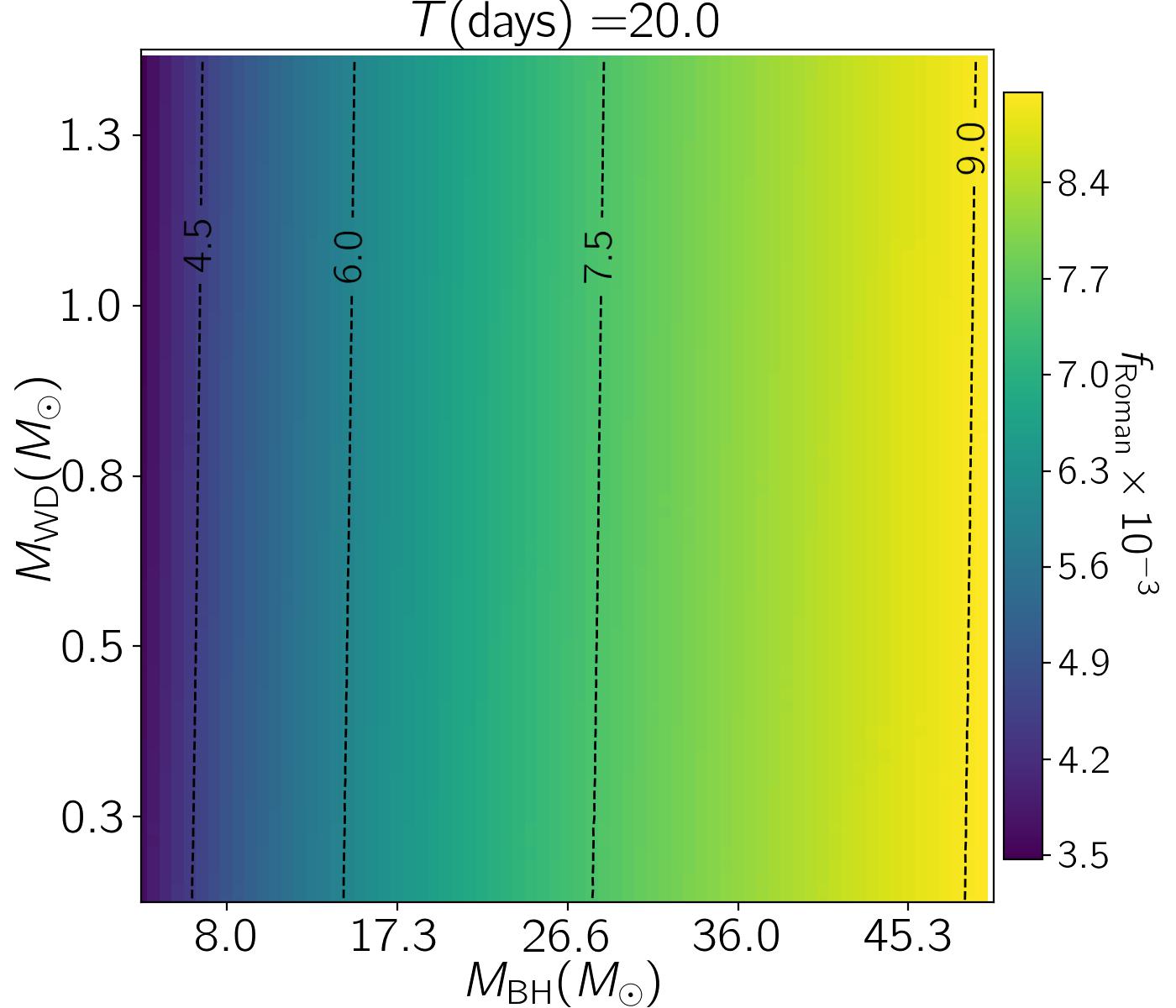}
\caption{Similar to the maps shown in Figure \ref{fig1}, but they display two statistical parameters. Left panels report $u_{0, m}$ the maximum lens impact parameter that offers SNR$=6$ (from Equation \ref{snr}) in WD+NS and WD+BH binary systems. Four other maps show the probabilities $f_{\rm{TESS}}$ and $f_{\rm{Roman}}$ which are the fractions of WD+NS and WD+BH systems offering self-lensing signals with SNR$\ge6$ by applying the TESS and Roman observing schedules, respectively.}
\label{fig1b}
\end{figure*}
\subsection{Evaluating Lensing Parameters in WD+NS and WD+BH systems}\label{sub22}
To identify the Einstein radius and normalized source radius for different WD+NS and WD+BH binary systems, we consider three free parameters including the masses of two components and their orbital periods. We choose these parameters from three wide ranges based on observations from these targets. The confirmed NSs (up to now) have the mass range $M_{\rm{NS}}\in[1.2,~2.4]M_{\odot}$ \citep{2018MNRASSuwa,2022ApJRomani}, while detected WDs have the mass range $M_{\rm{WD}}\in[0.17,~1.4]M_{\sun}$. From these two ranges, we uniformly select the masses of NSs and WDs. We also choose the mass of stellar-mass black holes from the range $M_{\rm{BH}}\in[3.3,~50]M_{\odot}$ smoothly. Their orbital periods could be from some minutes to some hundred days depending on their age and their formation channel. Therefore, we take their orbital periods discretely in the logarithmic scale and from the wide range $[2\rm{hrs},~50\rm{days}]$. 
We uniformly divide the two-dimensional (2D) space $M_{\rm l}-M_{\rm WD}$ into $10^4$ grids, and for each grid (with the specified masses of the WD and its companion as the lens object) we consider $50$ discrete values of the orbital periods, and calculate $R_{\rm E}$, $\rho_{\star}$, and the magnification factor $A(u, \rho_{\star})$ with $u=\rho_{\star}$. Here, $\rho_{\star}=R_{\rm WD}/R_{\rm E}$ is related to self-lensing signal of WDs by its companions, and $M_{\rm l}=M_{\rm NS}$ or $M_{\rm BH}$. Self-lensing signals of NSs by WDs are barely detectable in the overall flux received by the observer. Here, we fix the other parameters, which are the lens distance from the observer $D_{\rm{l}}=1$ kpc, the orbital eccentricity $\epsilon=0$, the inclination angle $i=0$, and $D_{\rm ls}=a$. We determine WDs' radius using the known mass-radius relation for WDs \citep{2017ParsonsMassrWD}.

In Figure \ref{fig1}, we represent six maps of $R_{\rm E}(R_{\oplus})$, $\rho_{\star}$, and the magnification factor $A$ for WD+NS (top row) and WD+BH (bottom row) binary systems over the 2D space $M_{\rm l}-M_{\rm WD}$ for $T=30$ days. Inside these maps the contour lines are displayed with dashed black curves. In right panels the counter lines of $\rho_{\star}$ are represented because $A$ strongly depends on this parameter. Similar maps related to other values of the orbital period are also available. According to the introduced lensing formalism, $R_{\rm E}$ and $\rho_{\star}$ as functions of the orbital period are:
\begin{eqnarray}
R_{\rm{E}}&=&2.7R_{\oplus}(\frac{M_{\rm l}}{1.4M_{\odot}})^{1/2}(\frac{M_{\rm l}+M_{\rm{WD}}}{2 M_{\odot}})^{1/6} (\frac{T}{30\rm{days}})^{1/3},\nonumber\\
\rho_{\star}&=&0.51\frac{R_{\rm{WD}}}{1.4R_{\oplus}}\frac{2.7R_{\oplus}}{R_{\rm E}}. 
\label{rero}
\end{eqnarray} 
Accordingly, $R_{\rm E}$ increases with the orbital period as $\propto T^{1/3}$, while $\rho_{\star}$ decreases by the orbital period as $\propto T^{-1/3}$. For close WD+NS systems $R_{\rm E}\sim R_{\oplus}$, which is similar to the radius of WDs. Hence, in these systems $\rho_{\star}\sim1$ (as shown in the middle panels of Figure \ref{fig1}), which affects on the magnification factor. Most of WD+NS binary systems with wide orbits and $T\gtrsim25$ days have $\rho_{\star}\lesssim1$, and those with $T\gtrsim150$ days have $\rho_{\star}\lesssim0.5$. Massive WDs (with $M_{\rm WD}\gtrsim1.3 M_{\sun}$) paired with NSs mostly have $\rho_{\star}\lesssim1$. Also, WD+NS systems with $M_{\rm{WD}}\gtrsim M_{\sun}$ when $T\gtrsim1$ day have $\rho_{\star}$ values less than one. 

\noindent For WD+BH binary systems, the three bottom panels of Figure \ref{fig1} show their $R_{\rm E}$, $\rho_{\star}$, and the magnification factor maps. By considering the adopted wide mass range for stellar-mass black holes, their $R_{\rm E}$ values are larger and vary over a wide range $\sim[0.01-1] R_{\sun}$. These systems have smaller values of $\rho_{\star}$, and higher magnification in comparison with WD+NS systems. WD+BH systems with $T\gtrsim3$ days have $\rho_{\star}$ less than one. For systems with $T\gtrsim 20$ day, their $\rho_{\star}$ values are less than $0.5$.  

\noindent Therefore, we expect that self-lensing signals in edge-on WD+BH systems have higher magnification with higher detectability rate in comparison with self-lensing signals in WD+NS systems. However, it has been estimated that the number of WD+BH systems will be smaller than the number of other types of compact binary systems (for examples, DWDs or WD+NS). The reasons for this are (1) the evolutionary timescales of BHs and WDs are different by three orders of magnitude, (2) the supernovae kick during BH formation will disturb their orbits, which can potentially disrupt the initial binary system, and (3) when the massive companion converts to a giant, an unstable mass-transfer likely occurs.  

\begin{figure*}
\centering
\includegraphics[width=0.45\textwidth]{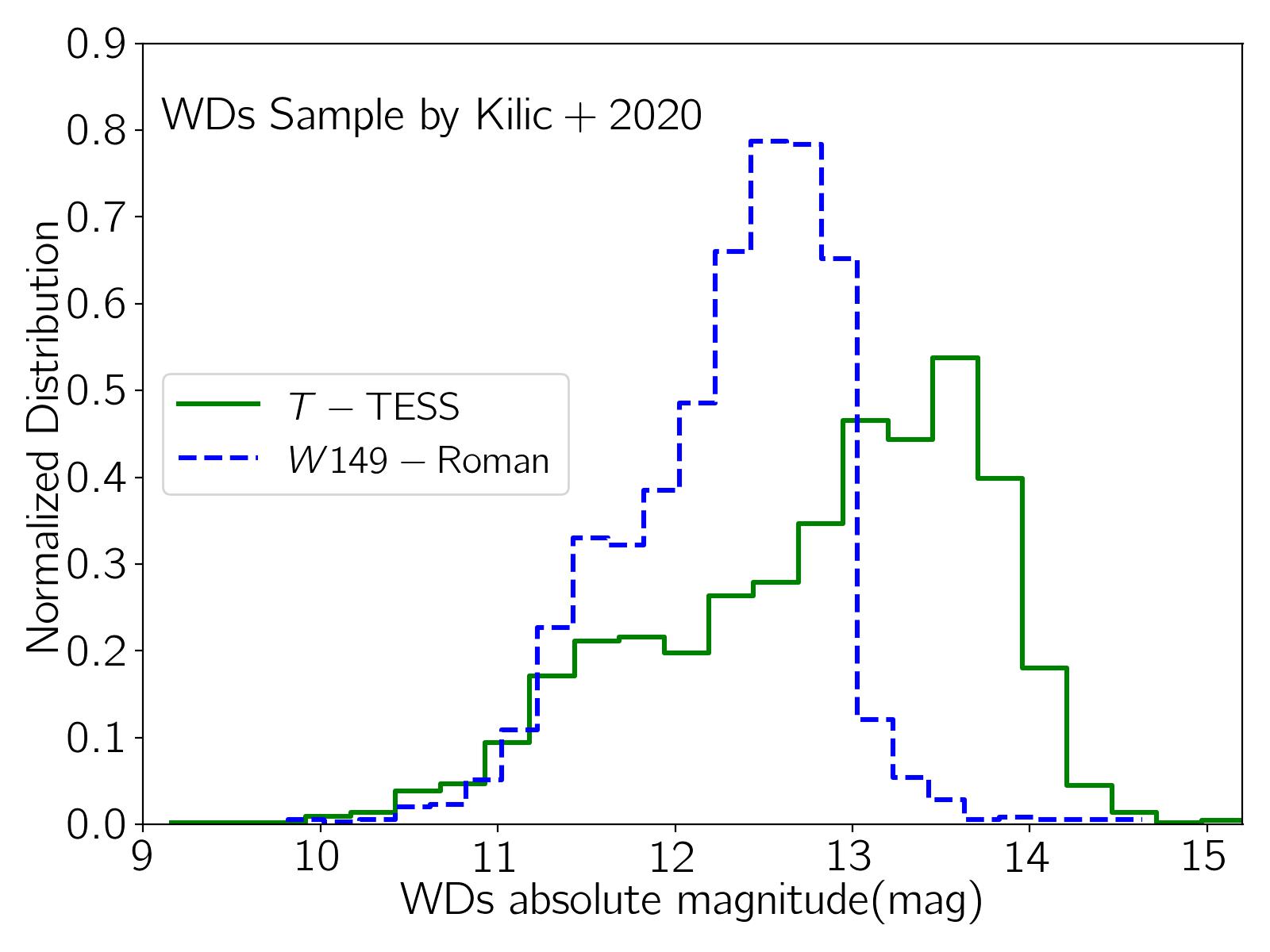}
\includegraphics[width=0.45\textwidth]{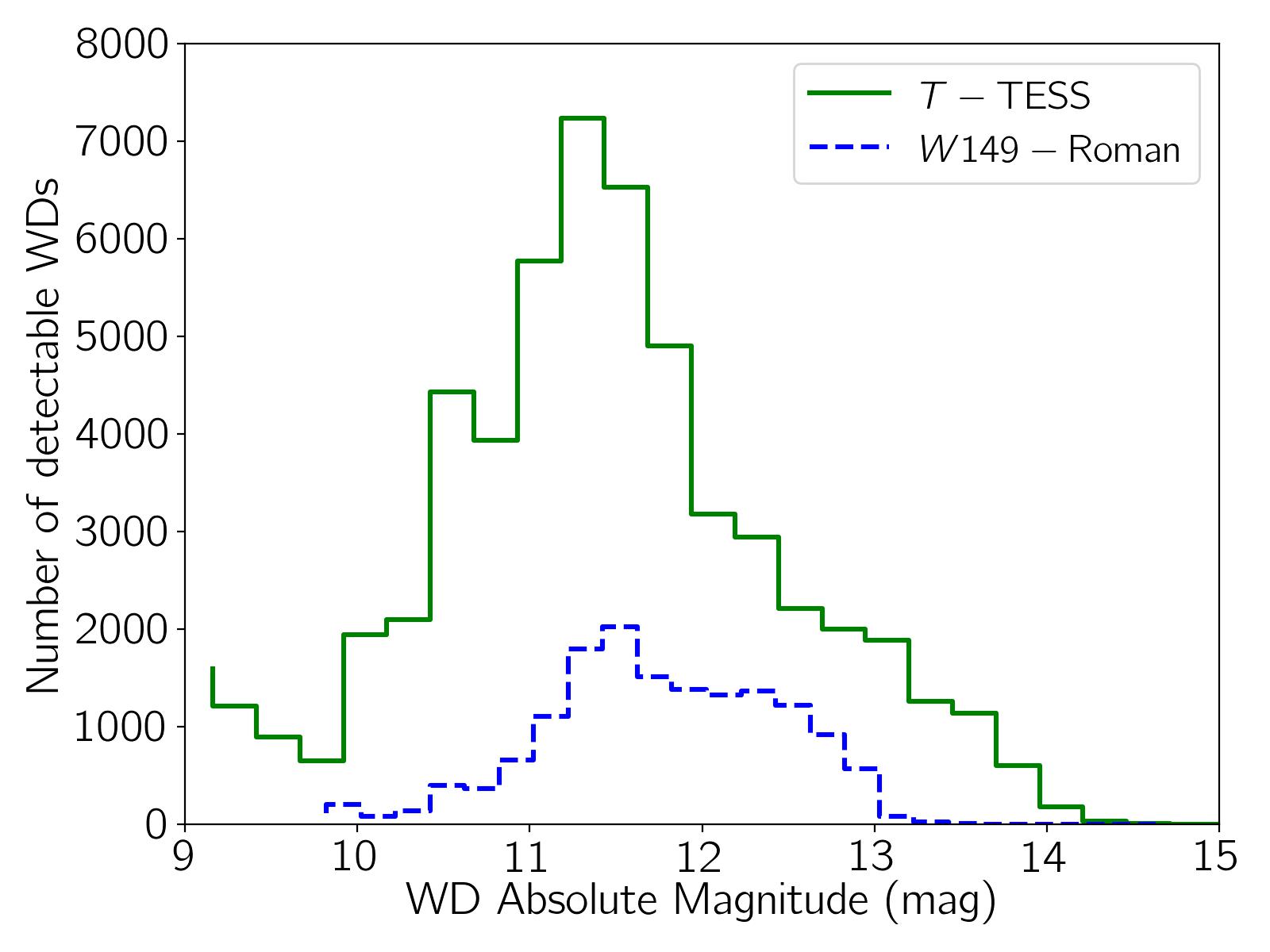}
\caption{Left panel: the normalized distributions of WDs offered by \citet{2020ApJkilic} versus their absolute magnitudes in the TESS $T$-band (solid green histogram), and the Roman W149 filter (dashed blue histogram). Right panel: the corresponding distributions of detectable WDs with the apparent magnitude in the TESS $T$-band $\leq17.5$ mag and in the Roman W149 $\in[14.8,~26]$ mag.}
\label{fig1c}
\end{figure*}
\subsection{Evaluating Statistical Parameters in WD+NS and WD+BH systems}\label{sub23}
Here, we aim to analytically evaluate the detectability of self-lensing signals in WD+NS and WD+BH systems. We first determine the maximum lens impact parameter $u_{0, m}$ so that the resulting self-lensing signal offers an enough signal to noise (SNR) value for detection. The SNR value for periodic self-lensing signals is given by: 
\begin{eqnarray}
\rm{SNR}=\sqrt{N_{\rm{tran}}}~\frac{A(u_{0, m}, \rho_{\star})-1}{\sigma_{\rm{A}}},
\label{snr}
\end{eqnarray} 
where, $N_{\rm{tran}}=T_{\rm{obs}}/T$, $T_{\rm{obs}}$ is the observing time, and $\sigma_{\rm{A}}$ is the photometric accuracy which depends on the stellar apparent magnitude. $T_{\rm{obs}}$ and $\sigma_{\rm{A}}$ depend on the applied observing strategy. By demanding SNR$\ge6$ we determine the maximum lens impact parameter $u_{0, m}$ based on Equation \ref{snr}.

\noindent Then, by having $u_{0, m}$ we determine the probability of detecting lensing signals using $f=\rm{Max}(u_{0, m}R_{\rm E}, R_{\rm{WD}})/a$, where the output of the function $\rm{Max}()$ is its largest argument. If $u_{0, m}R_{\rm E}\leq R_{\rm{WD}}$, eclipsing happens in addition to self-lensing signals. The probability of $f$ is the fraction of all possible positions of the WD around its companion in a circular orbit (i.e., $4\pi a^{2}$) in which self-lensing signals happen with SNR$\ge6$.

To evaluate the probability $f$, we apply two observing strategies by the TESS and Roman telescopes, and set $T_{\rm{obs}}=27.4,~62$ days, and $\sigma_{\rm A}=0.007,~0.001$ for these observations, respectively. The applied photometric accuracies are related to stellar apparent magnitudes $\sim15$ mag. In Figure \ref{fig1b}, we show the maps of $u_{0, m}$ and the probability $f$ by applying observing strategies by the TESS ($f_{\rm{TESS}}$) and Roman ($f_{\rm{Roman}}$) telescopes. According to Equation \ref{rero}, the Einstein radius enhances with the lens mass as $R_{\rm{E}}\propto M_{\rm l}^{2/3}$. Hence, $u_{0, m}$ maximizes for massive WD+NS and WD+BH binary systems. We note that massive WDs have smaller radii ($R_{\rm{WD}}\propto M_{\rm{WD}}^{-1/3}$) and as a result smaller $\rho_{\star}$ values. Two left maps in Figure \ref{fig1b} confirm this point. For a given orbital period, the semi-major axis is $a\propto(M_{\star}+M_{\rm l})^{1/3}$, i.e., more massive binary systems are wider when the orbital period is fixed. Hence, $f$ maximizes for WD+NS and WD+BH binary systems with low-mass WDs revolving massive NS/BH objects. Generally, these probabilities are $\sim10^{-3},~10^{-2}$ for WD+NS and WD+BH binary systems, respectively.

To estimate the number of detectable self-lensing signals, we evaluate the number of detectable WDs during the TESS and Roman Galactic Exoplanet Survey (RGES) in the observing filter $F$ using:
\begin{eqnarray}
N_{\rm{WD}, F}=\int_{M_{F}}\xi_{\rm{WD}}dM_{F}\int_{V(M_{F})}\frac{\rho_{\rm{WD}}(R, z)}{\overline{M}_{\rm{WD}}}~dV,
\label{nwd}
\end{eqnarray}
where, $M_{F}$ is the absolute magnitude of WDs in the applied filter $F$, $\rho_{\rm{wd}}(R, z)$ is the Galactic spatial density distribution of WDs which is a function of $R$ (the radial distance from the Galactic center) and $z$ (the vertical distance from the Galactic plane). $\overline{M}_{\rm{WD}}$ is the average mass of WDs, and $\xi_{\rm{WD}}(M_{F})$ is the WD number density versus their absolute magnitude. In Equation \ref{nwd}, there are two integrals over (1) the WD absolute magnitude, and (2) the volume $V(M_{F})$ in which WDs with the given absolute magnitude $M_{F}$ are detectable. We note that the maximum distance from the Sun in which the WDs with the absolute magnitude $M_{F}$ are detectable by ignoring the extinction effect is given by $D_{\rm{max}, F}= 10\rm{pc}\times10^{0.2(m_{\rm{th}, F}-M_{F})}$  which offers the WD apparent magnitude less than the detection threshold $m_{\rm{th}, F}$. To convert the observer coordinate $(D, l, b)$ to the Galactic coordinate $(R, z)$ we use the relations given in Appendix of \citet{2023AJSahuSajadian}. Here, $(l, b)$ are the Galactic longitude and latitude, respectively.

To estimate $N_{\rm{WD}, F}$ in Equation \ref{nwd}, we take the WDs from a real ensemble containing $1772$ confirmed WDs within $100$ pc through the SDSS observations which was offered by \citet{2020ApJkilic}. From this sample, we get $\overline{M}_{\rm{WD}}=0.66M_{\odot}$. In this sample, the apparent magnitudes of WDs were given in the Gaia passbands. Based on their parallax amounts, and by estimating the extinction using $A_{v}(\rm{mag})=0.7 D(\rm{kpc})$, we derive their absolute magnitudes in the Gaia filters. Then we calculate their absolute magnitudes in the TESS $T$ and Roman $W149$ filters using their known relations \citep{2008AJLasker,2015ApJSAlam,2018AJStassun}. In the left panel of Figure \ref{fig1c}, we show the normalized distributions of the WDs' absolute magnitude in the TESS $T$-band (solid green) and Roman $W149$ (dashed blue). These distributions represent $\xi(M_{F})$. However, they should be normalized so that $\int \xi(M_{F})~dM_{F}=1$. We take WDs spatial density distribution $\rho_{\rm{WD}}(R, z)$ from Besan\c{c}on model related to the Galactic thin disk. In this model, the local number density of WDs is $\sim0.01\rm{pc}^{-3}$. By calculating integrals in Equation \ref{nwd}, we estimate the number of WDs detectable in the TESS and RGES observations versus their absolute magnitudes ($dN_{\rm{WD}, F}/dM_{F}$) as plotted in the right panel of Figure \ref{fig1c}. The areas under these curves represent $N_{\rm{WD}, F}$. By considering $m_{\rm{th}, T}=17.5$ mag, we get $N_{\rm{WD}, T}\simeq55,600$. For the RGES observations, we limit the observing area to $2~\rm{deg}^{2}$ within the Galactic thin disk, and set $m_{\rm{th}, W149}=26$ mag which results $N_{\rm{WD}, W149}\simeq15,000$. We note that it was predicted that during Roman High Latitude Survey (HLS) around $735,000$ WDs would be discovered but its observing cadence ($~120$ days) is too long to discover self-lensing signals \citep{2020ApJfantin}.

Accordingly, we roughly estimate the number of detectable self-lensing signals in WD+NS and WD+BH binary systems as $f\times f_{\rm b}\times N_{\rm{WD}}\sim10-50\times f_{\rm b}, 100-500\times f_{\rm b}$ respectively, where we assume the fraction $f_{\rm b}$ of WDs have NS or BH companions. However, in reality the detectability of these signals strongly depend on the orbital period, WDs' apparent brightness, the lens mass, the photometric errors, and the observing cadence, whereas none of them were considered in these rough estimations. Additionally, to accurately estimate the number of detectable self-lensing signals, we should generate these systems based on their realistic distributions.

In the next section, we construct two samples of possible light curves generated by different WD+NS and WD+BH systems by adopting the related parameters from their known distributions and study their detectability in the TESS and Roman survey observations. \\

\section{WD+NS and WD+BH Binary Systems: Simulations and Results}\label{sec3}
In this section, we construct realistic Monte Carlo simulations from WD+NS and WD+BH binary systems, as explained in Subsection \ref{sub31} by details. We then generate their light curves and synthetic data points by applying two observing strategies as described in Subsection \ref{sub32}. We finally evaluate detectability of their self-lensing signals and offer the statistical results in Subsection \ref{sub33}.

\subsection{Monte-Carlo simulations}\label{sub31}
{\bf Generating WDs population}:~We use the catalog of WDs offered by \citet{2020ApJkilic}, and assume that these WDs have NS or BH companions. For the WDs in this ensemble, their physical parameters (including mass, radius, distance, surface temperature, surface gravity, galactic coordinate), and their apparent magnitudes in the Gaia bands ($G$, $G_{\rm{BP}}$, and $G_{\rm{RP}}$) were reported. We note that WDs farther than $100$ pc can be detectable through the TESS and Roman observations. 

In our simulations, we determine the WD distance from the observer ($D_{\rm l}$) based on the number density versus distance, i.e., $dN/dD_{\rm l}\propto\rho(D_{\rm l})D_{\rm l}^{2}$. Here, $\rho(D_{\rm l})$ is the overall spatial density in our galaxy and we take its components from the Besan\c{c}on model \citep{robin2003,robin2012}. In the TESS observation, WDs up to $D_{\rm{max}, T}\sim500$ pc are detectable based on the TESS detection threshold $m_{\rm{th}, T}=17.5$ mag. During RGES, the Roman telescope covers a field with the projected area $\sim2~\rm{deg}^{2}$ toward the Galactic bulge. 

To calculate their photometric errors during observations by the TESS and Roman telescopes, we first transform their absolute magnitudes to those in the TESS $T$-band and the standard Johnson-Cousins filters as explained in Section (3) of \citet{2025AJsajadianmicrotess} and then calculate their apparent magnitudes. For WDs, we indicate their linear limb-darkening coefficients according to their effective surface temperature and their surface gravity \citep{2020AAclaretLimb}.

{\bf Generating NSs/BHs populations:} We determine the mass of NSs from a Gaussian distribution with the mean and width equal to $1.28~M_{\sun}$ and $0.24~M_{\sun}$, respectively \citep{2012ApJOzel}. We specify NSs' radii according to their masses \citep{2019UnivLattimer}. The age of NSs in our simulation is selected from a log-uniform distribution over the range $\log_{10}[\rm{Age}(\rm{yrs})]\in[4, 8]$. We choose the mass of BHs uniformly from the range $[3.3,~50]M_{\odot}$. 

To generate binary orbits, we first determine their semi-major axis $a$ from the \"{O}pik's law \citep{1924Opik}, i.e., a log-uniform distribution, within a wide range $\log_{10}[a(R_{\rm{WD}})]\in[0.5,~6]$, here $R_{\rm{WD}}$ is the WD' radius. However, we exclude contact binaries whose Roche lobe is less than the source radius. The orbital eccentricity is selected uniformly over the range $[0, \epsilon_{\rm{max}}]$, where the upper limit $\epsilon_{\rm{max}}$ is determined using the known eccentricity-period relation \citep[see, e.g., ][]{2008EASMazeh}. The projection angles $i$ and $\theta$ are selected uniformly over the ranges $[0,~90]$ and $[0,~360]$ degree, respectively. However, self-lensing signals due to binary orbits that are not edge-on and with larger inclination angles are very unlikely to be detected. $t_{\rm p}$ is uniformly selected over the range $[0,~T]$ days.      

To include the blending effect due to NS companions (although it is very small) in the overall flux, we need the ratio of brightness of NS to brightness of WD in a given passband, i.e., $\mathcal{F}$. We calculate this ratio numerically by integrating over their Planck distributions in the wavelength range of applied filters (i.e., the TESS $T$-band and the Roman W149) by considering their throughput functions as the weight functions. The occultation effect due to NSs during self-lensing of WDs is ignorable. When NSs' brightness is considerable and while its WD companion is passing in front of it, the WD will block its brightness completely, and a complete eclipsing happens.

\begin{figure*}
    \centering
    \includegraphics[width=0.49\textwidth]{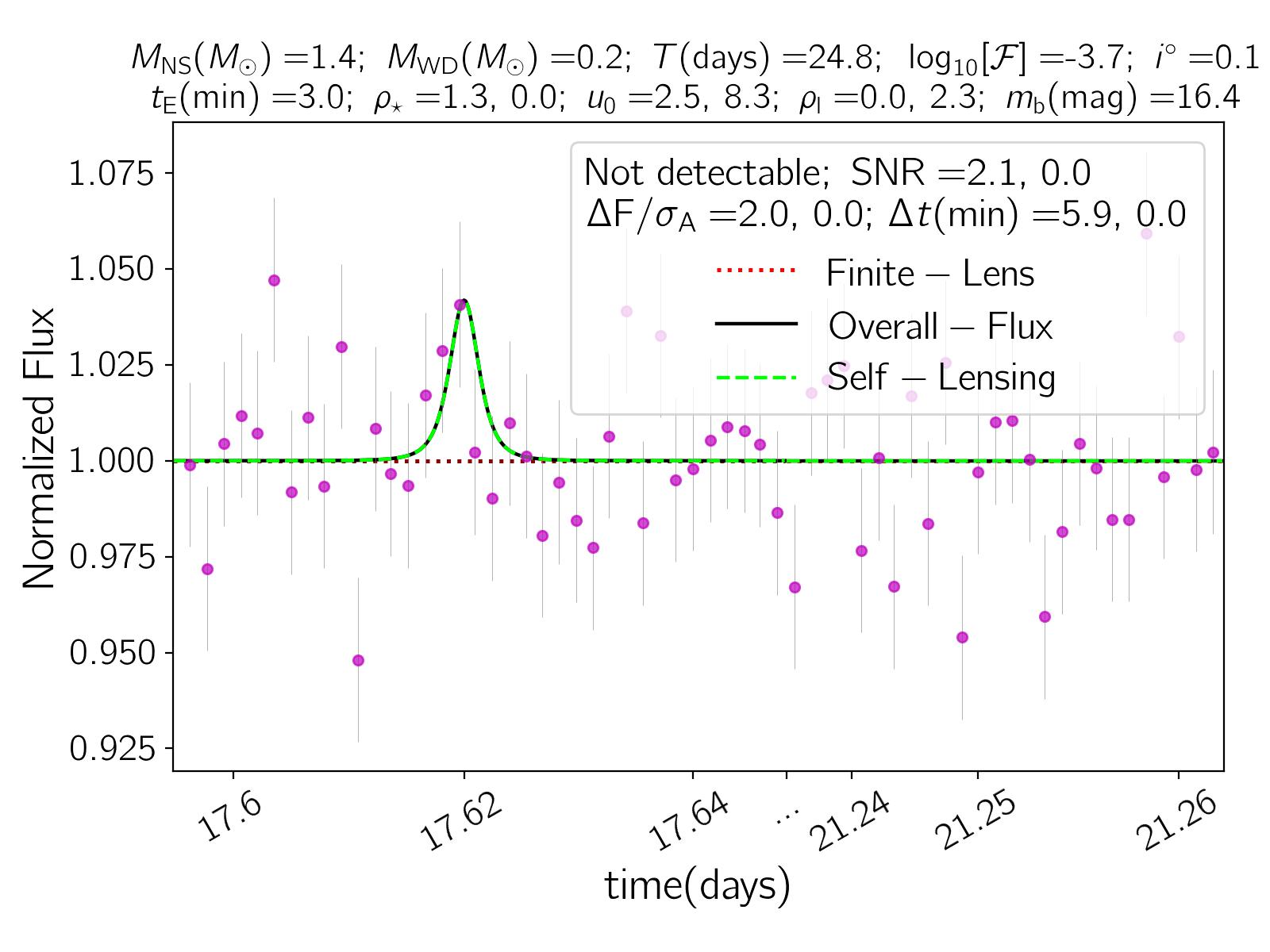}
    \includegraphics[width=0.49\textwidth]{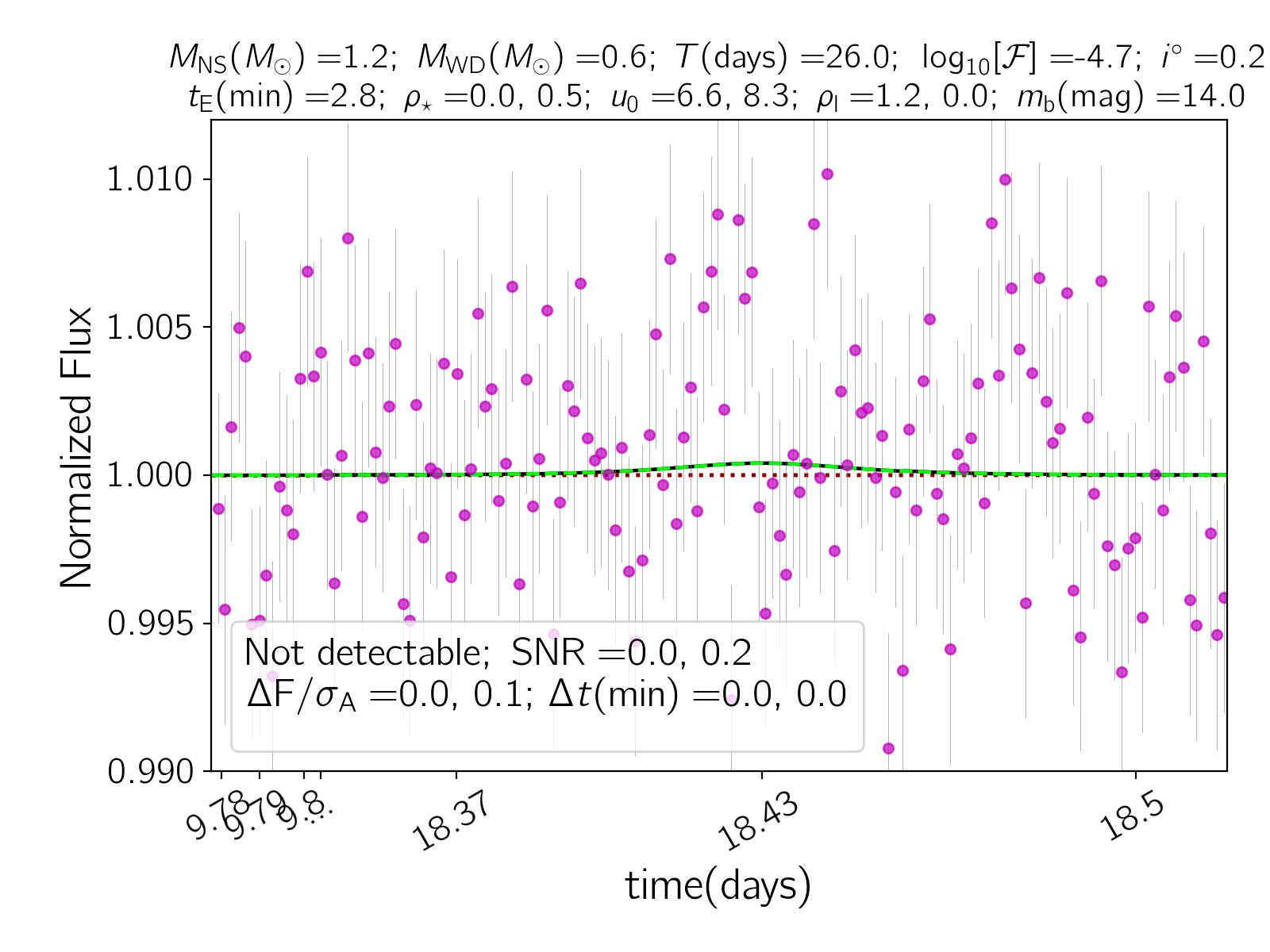}
    \includegraphics[width=0.49\textwidth]{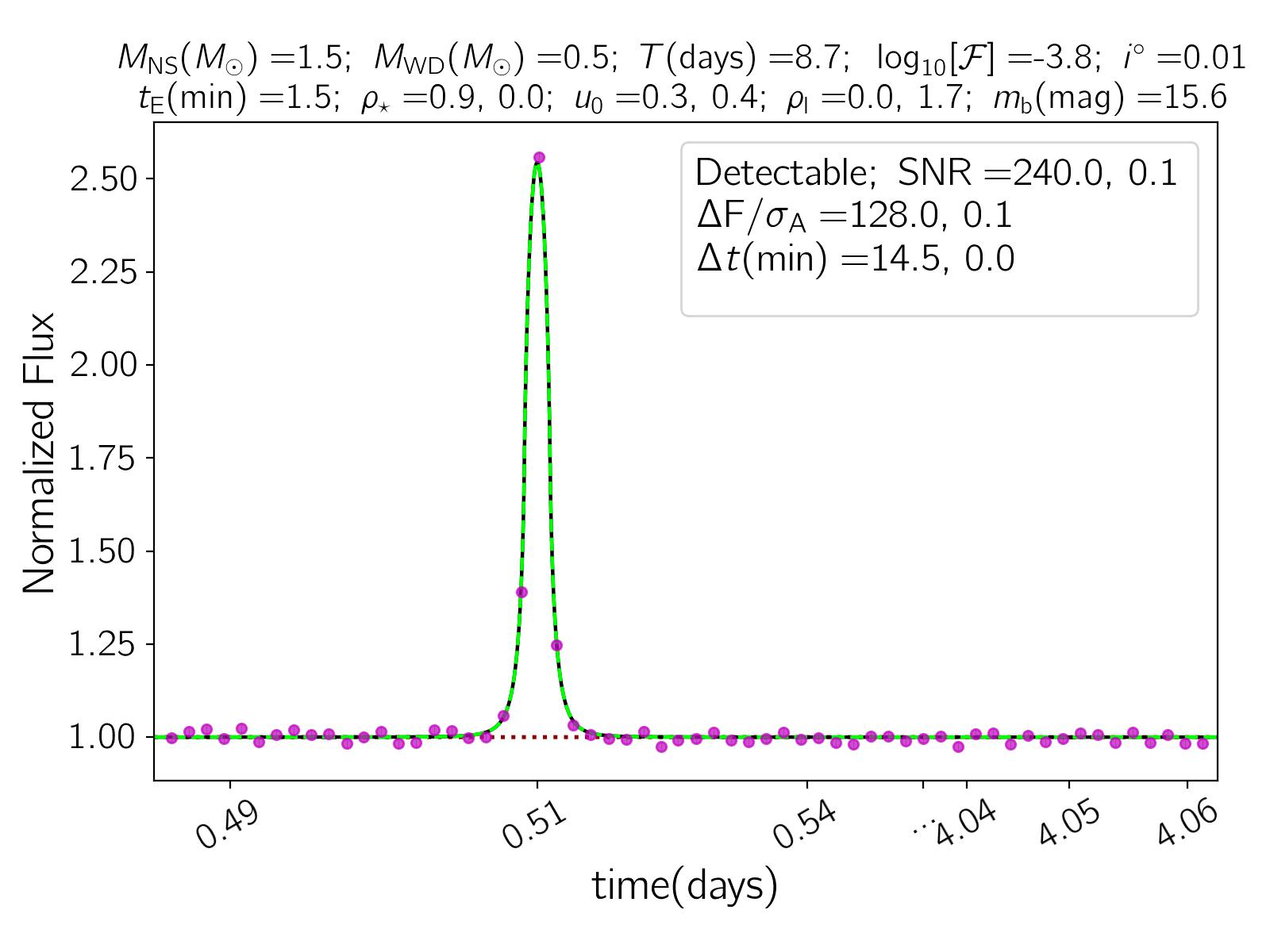}
    \includegraphics[width=0.49\textwidth]{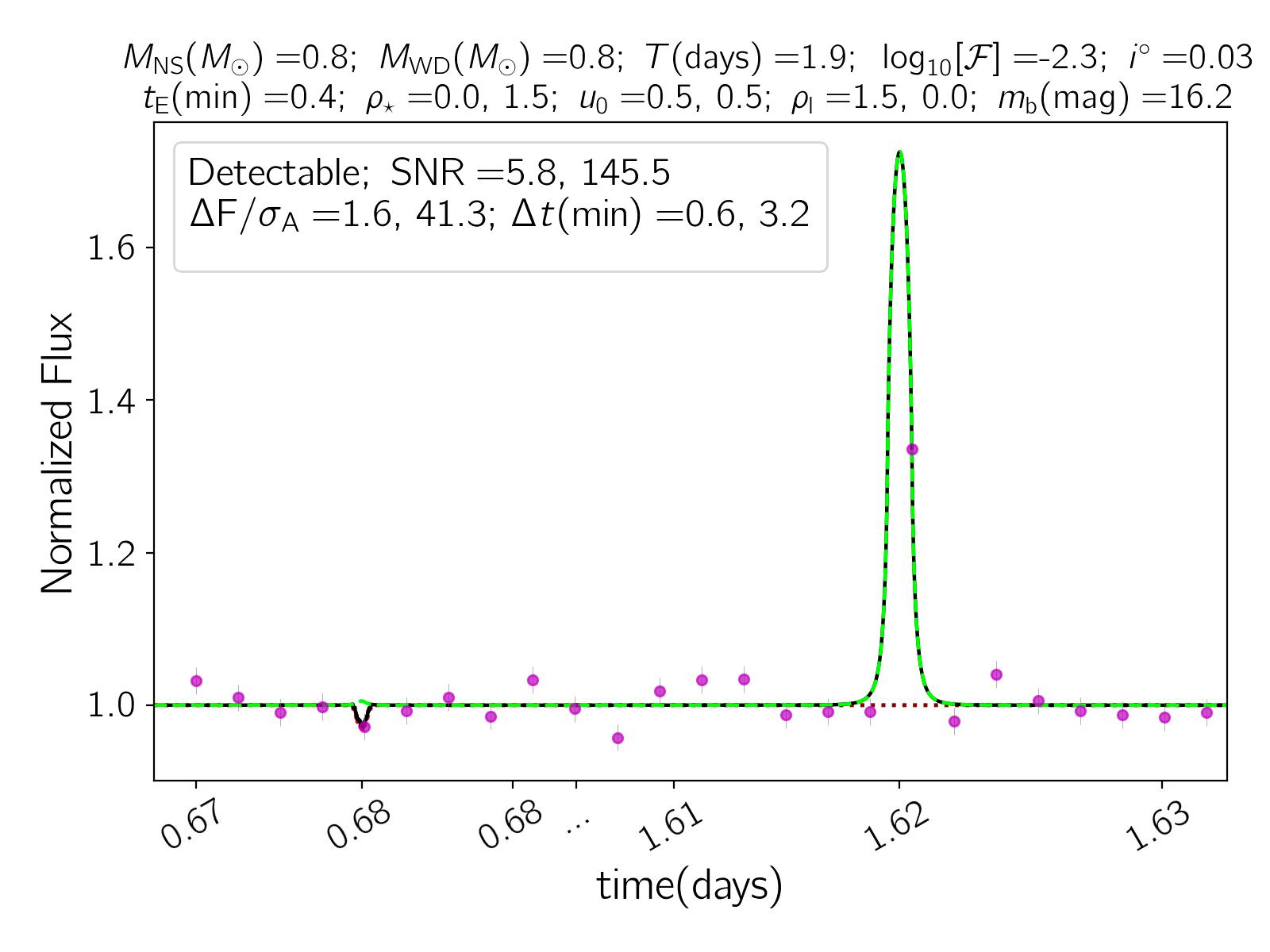}
    \includegraphics[width=0.49\textwidth]{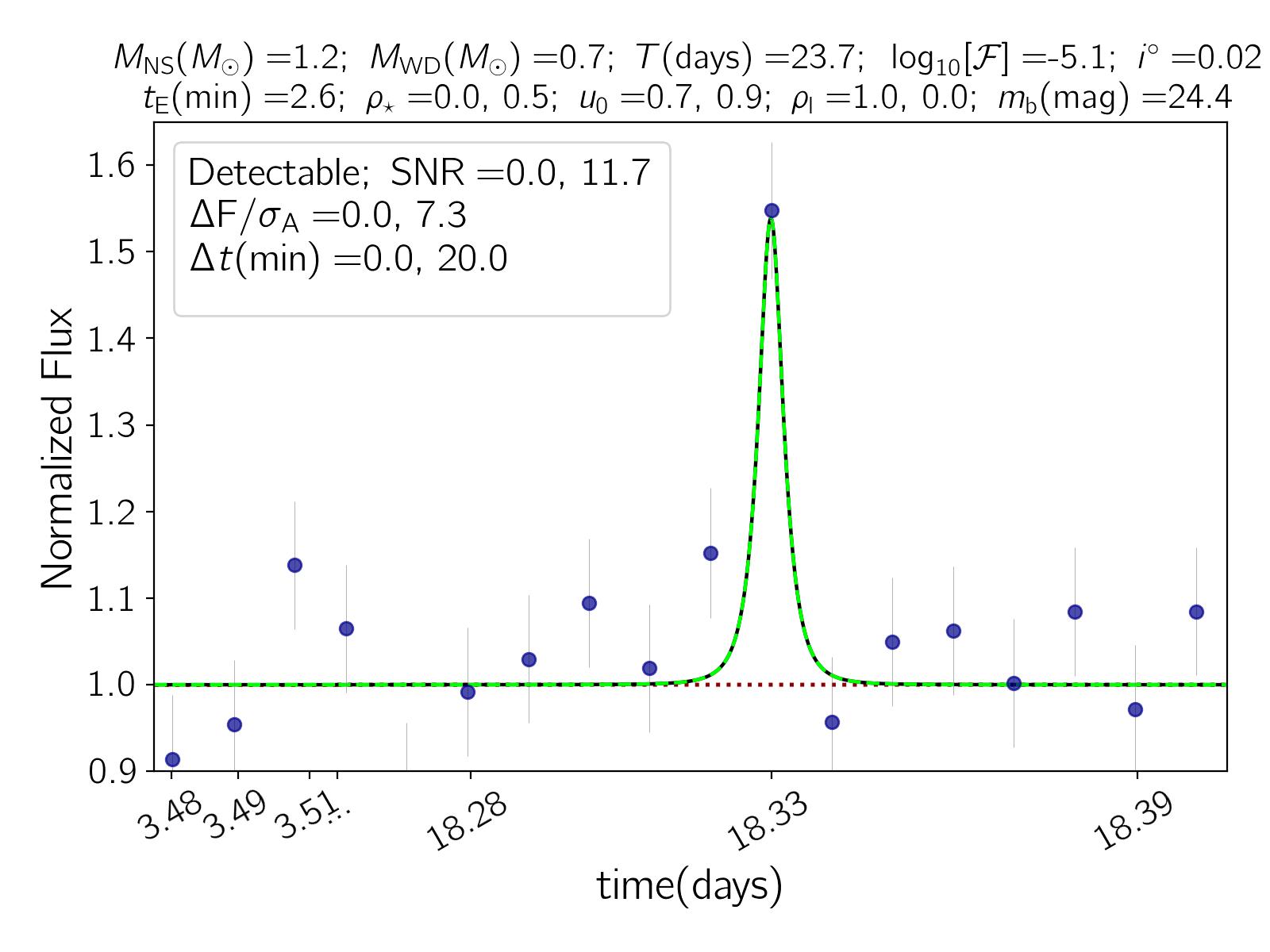}
    \includegraphics[width=0.49\textwidth]{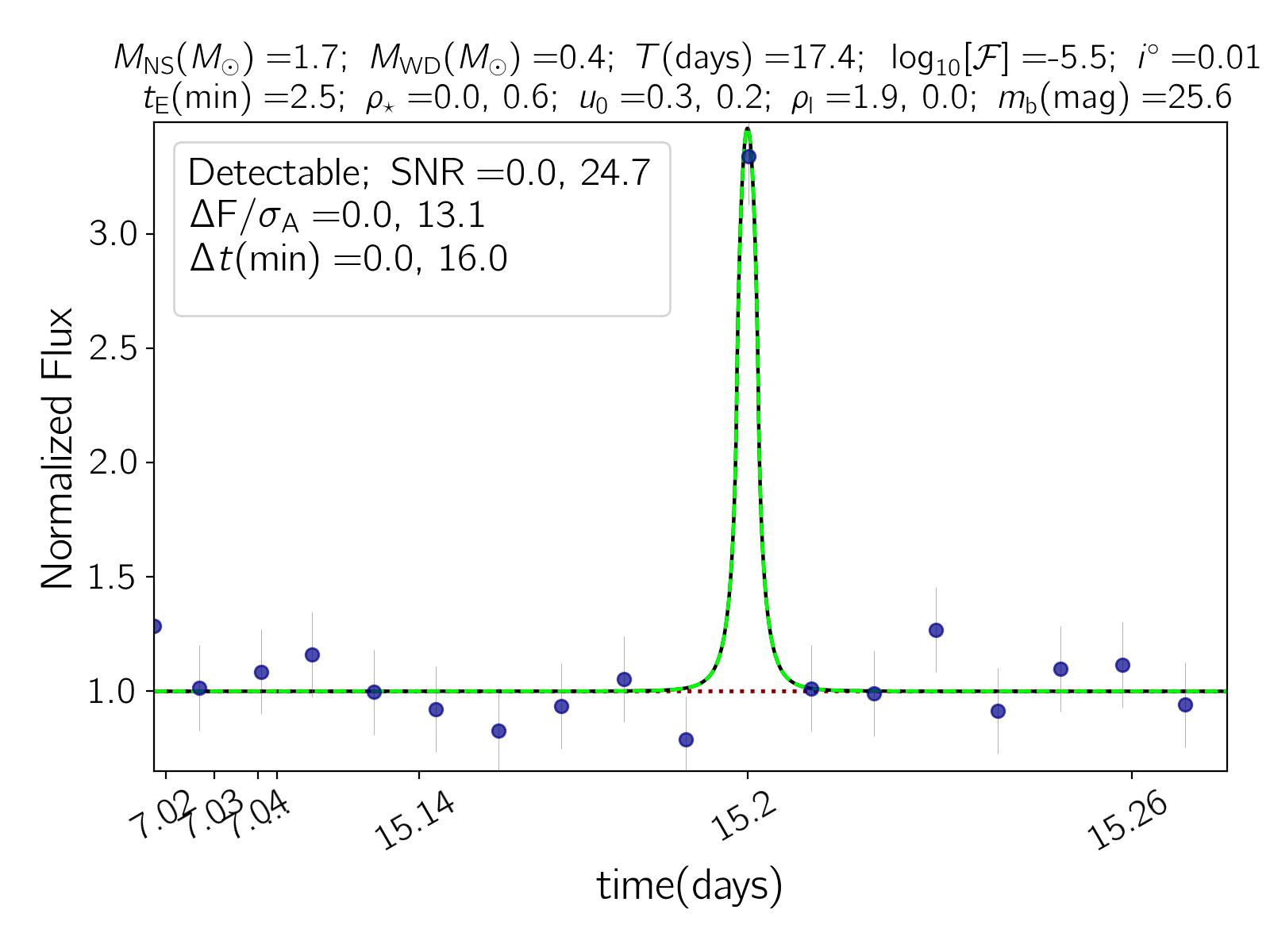}
    \caption{Six examples of simulated light curves due to WD+NS binary systems. The solid black, dotted red, and dashed lime curves represent the overall normalized flux by considering both self-lensing and eclipsing effects, i.e., $A_{\rm{tot}}$ which is given by Eq. \ref{atot}, the contributions of occultation and self-lensing effects in the overall flux, respectively. The adopted physical parameters (top row) and resulting lensing parameters (bottom row) are mentioned at top of plots. The properties of signals (related to their detectability) are mentioned in the box inside each panel. The horizontal axes show two zoomed parts while passing two components in front of each other. The synthetic magenta (inside four top panels) and  dark blue (within two last panels) data points are taken by the TESS and Roman telescopes, respectively.}
    \label{fig2}
\end{figure*}

In the following subsection, we explain making light curves due to WD+NS and WD+BH binary systems by applying two observing strategies, and offer some examples of simulated light curves.

\subsection{Applied Observing strategies}\label{sub32}
To simulate synthetic data points and extract light curves with detectable self-lensing signals, we apply two observing strategies, which are described as follows. 

{\bf TESS observations}: The TESS observing time span for each sector is $27.4$ days. Although these sectors have some overlapped area, $\sim 74\%$ of stars are detectable by TESS from one sector and accordingly observed continuously during $27.4$ days (with a one-day gap in middle) every two years. Hence, we simulate synthetic data points taken by TESS during $27.4$ days. We determine the TESS photometric error in the magnitude unit ($\sigma_{\rm{m}}$), which for the TESS and Kepler observations is usually expressed in the form of the Combined Differential Photometric Precision (CDPP) metric, according to the apparent magnitude of stars in the TESS $T$-band ($m_{\rm{T}}$). Using stellar catalogs extracted by the TESS-SPOC pipeline \citep{2020RNAAStessspoc}, we derive a linear curve for $\sigma_{\rm{m}}$ versus $m_{\rm{T}}$ for stars detected by TESS \citep[one example can be found in Fig. 1 of ][]{2025AJsajadianmicrotess}. We fix the TESS observing cadence to $\tau=2$ minutes, which was used for observing the TESS Candidate Target List (CTL, \citet{2018AJStassun}). 

{\bf Roman observations}: The Roman telescope will observe the Galactic bulge during six $62$-day seasons with a $15$-min cadence. This survey observation, the so-called Roman Galactic Exoplanet Survey (RGES), is dedicated to detect microlensing events due to exoplanet, both bound and free-floating \citep{2013Spergelwfirst,2019ApJSPenny,2020AJJohnson}. Nevertheless, it has been predicted that many other transient astrophysical phenomena would be discovered during RGES \citep[e.g., ][]{2023AJfathmulti,2023MNRASmoonplanet,2025MNRASbffs}. In this work, we study detecting self-lensing signals from WD+NS binary systems during one of its $62$-day seasons. The Roman photometric accuracy also depends on stellar apparent magnitudes in the Roman filter \citep{2019ApJSPenny}.  

In Figure \ref{fig2}, we show six simulated light curves due to different edge-on WD+NS binary systems. The solid black curves show the overall flux due to both components normalized to the baseline, as: 
\begin{eqnarray}
A_{\rm{tot}}=\frac{1}{1+\mathcal{F}}\Big[A_{\rm{WD}}-\mathcal{O}_{\rm{WD}}+\mathcal{F}\times \big(A_{\rm{NS}}-\mathcal{O}_{\rm{NS}}\big)\Big], 
\label{atot}
\end{eqnarray}
where $A$ represents the magnification factor during self-lensing, $\mathcal{O}$ shows the occultation effect and equals to the blocked portion of images' area by the lens normalized to the un-lensed source area. The first term $A_{\rm{WD}}-\mathcal{O}_{\rm{WD}}$ is the net magnification factor for the WD's brightness due to passing behind its NS companion, which is a function of time $t$. The second term $A_{\rm{NS}}-\mathcal{O}_{\rm{NS}}$ represents the net variation in the NS's brightness due to passing behind its WD companion at a given time $t$. The second term has a negligible contribution in $A_{\rm{tot}}$ because of the smallness of $\mathcal{F}$. The dotted red curves display the occultation contribution in the overall flux, i.e., $1-\big(\mathcal{O}_{\rm{WD}}+\mathcal{F}\times\mathcal{O}_{\rm{NS}}\big)/(1+\mathcal{F})$. The dashed lime curves display the self-lensing contribution to the overall flux $\big(A_{\rm{WD}}+\mathcal{F}\times A_{\rm{Ns}}\big)/(1+\mathcal{F})$. 

\noindent The physical (top row) and lensing (bottom row) parameters due to each light curve can be found at its top. $m_{\rm{b}}$ represents the baseline apparent magnitude in the applied filter ($T$ and W149 for the TESS and Roman observations, respectively). We estimate the Einstein crossing time for these self-lensing signal as $t_{\rm{E}}=R_{\rm{E}}/v_{c}$, where $v_{c}=2\pi a/T$ is the velocity in a circular and completely edge-on orbit with the radius equals to the semi-major axis $a$. In elliptical orbits the velocity is a function of time and variable.

\noindent Inside the legend of each panel, we report the properties of signals including SNR (given by Eq.  \ref{snr}), $\Delta F/\sigma_{\rm A}$, and $\Delta t$. The second one is the signal's depth ($\Delta F=|A_{\rm{tot}, max}-1|$, where $A_{\rm{tot}, max}$ is the maximum value of the overall normalized flux) which is divided by the photometric error, i.e., $\sigma_{\rm{A}}=\big|1-10^{-0.4 \sigma_{\rm{m}}}\big|$. Also, $\Delta t$ represents the signal's duration and we calculate it numerically. In simulations, we adopt the time interval in which $A_{\rm{tot}}\geq 1+\sigma_{\rm{A}}/2$ (whenever the overall magnification exceeds from half of the photometric error) as the signal's duration ($\Delta t$). This duration is mostly longer than the Einstein time scale $t_{\rm E}$ because it does not represent the signal's width. One can find this point from panels of Figure \ref{fig2}.

\noindent For some parameters ($\Delta t$, $\Delta F/\sigma_{\rm A}$, $\rho_{\star}$, $\rho_{\rm l}$, and $u_{0}$ the lens impact parameter) two values are given, related to the first and second lensing/eclipsing signal in the chronological order. We note that the horizontal axes only show two zoomed parts while passing each component in front of the other component. The simulated synthetic data points taken by the TESS and Roman telescopes are displayed by magenta and dark blue colors, respectively.

Self-lensing/eclipsing signals in the two top light curves are not recognizable because the amplitudes of the signals are less than or similar to the photometric errors of data. The four remaining light curves have detectable self-lensing/eclipsing signals because their inclination angles are small: $\sim 0.01-0.03$ degree. In the fourth light curve, the eclipse of the NS's flux due to passing behind its WD companion is somewhat recognizable because for this target $\mathcal{F}\simeq0.005$. We generate synthetic data points taken by Roman for the two bottom light curves with a $15$-min cadence, which yields substantially lower time sampling.

\noindent Accordingly, detection of self-lensing/eclipsing signals in WD+NS systems through the TESS or Roman observations needs a tight alignment between two components with $i\lesssim 0.1$ degree. From simulations, we find that the average duration of simulated signals (due to WD+NS systems with $T\leq 27.4$ days) is $\Delta t \sim6$ minutes. Even with the $2$-min observing cadence, only a few data points can be recorded during self-lensing/eclipsing signals, which is confirmed by the  light curves shown in Figure \ref{fig2}. Also, when the WD is passing in front of its NS companion, although $\rho_{\star}$ is very small (yielding a higher magnification), and $\rho_{\rm l}\sim 1$ (which can generate a partial or complete eclipse), this variation in the overall flux is barely recognizable because of the negligible contribution of NSs (i.e., $\mathcal{F}\ll1$). 
\begin{figure}
    \centering
    \includegraphics[width=0.49\textwidth]{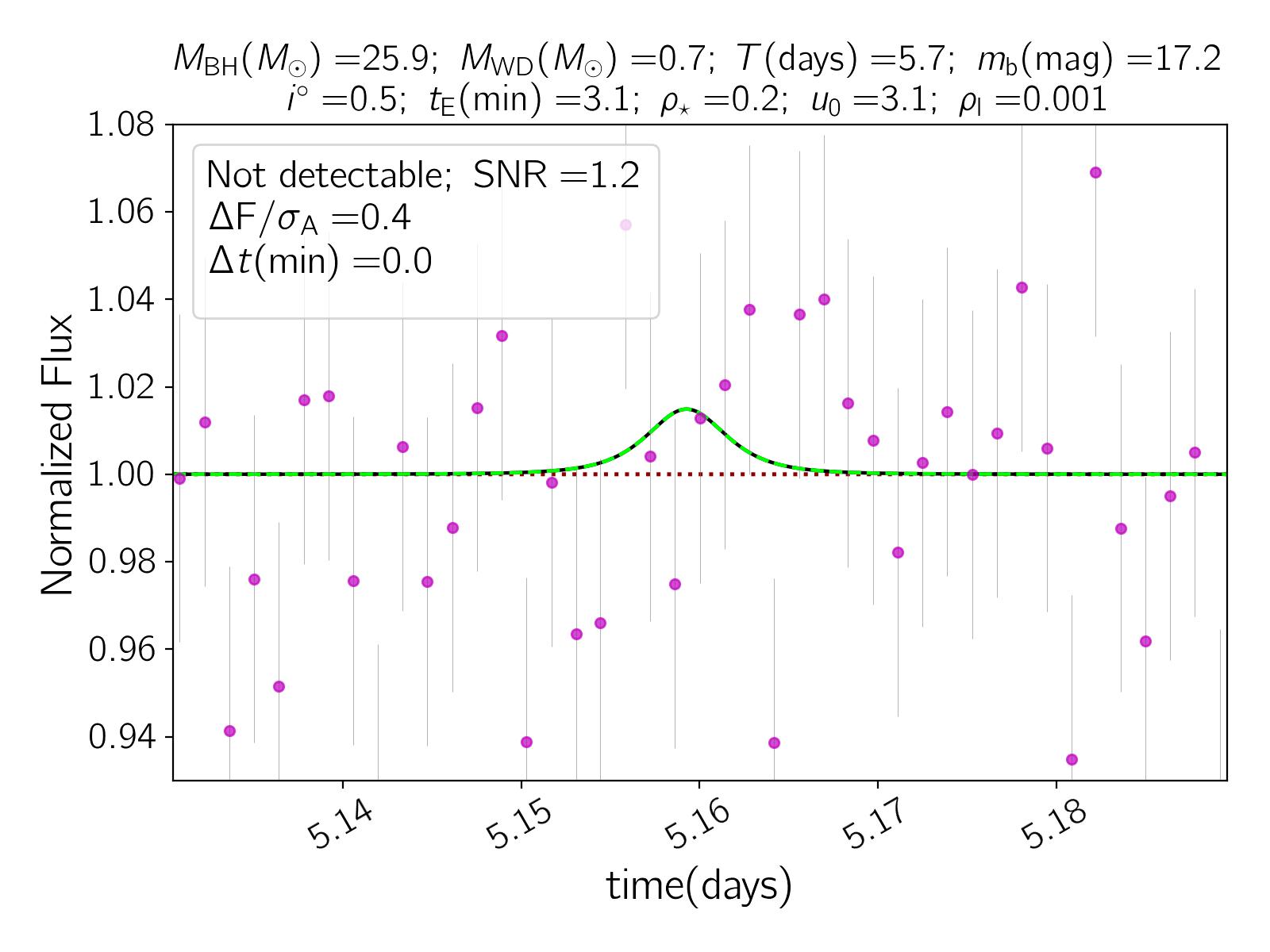}
    \includegraphics[width=0.49\textwidth]{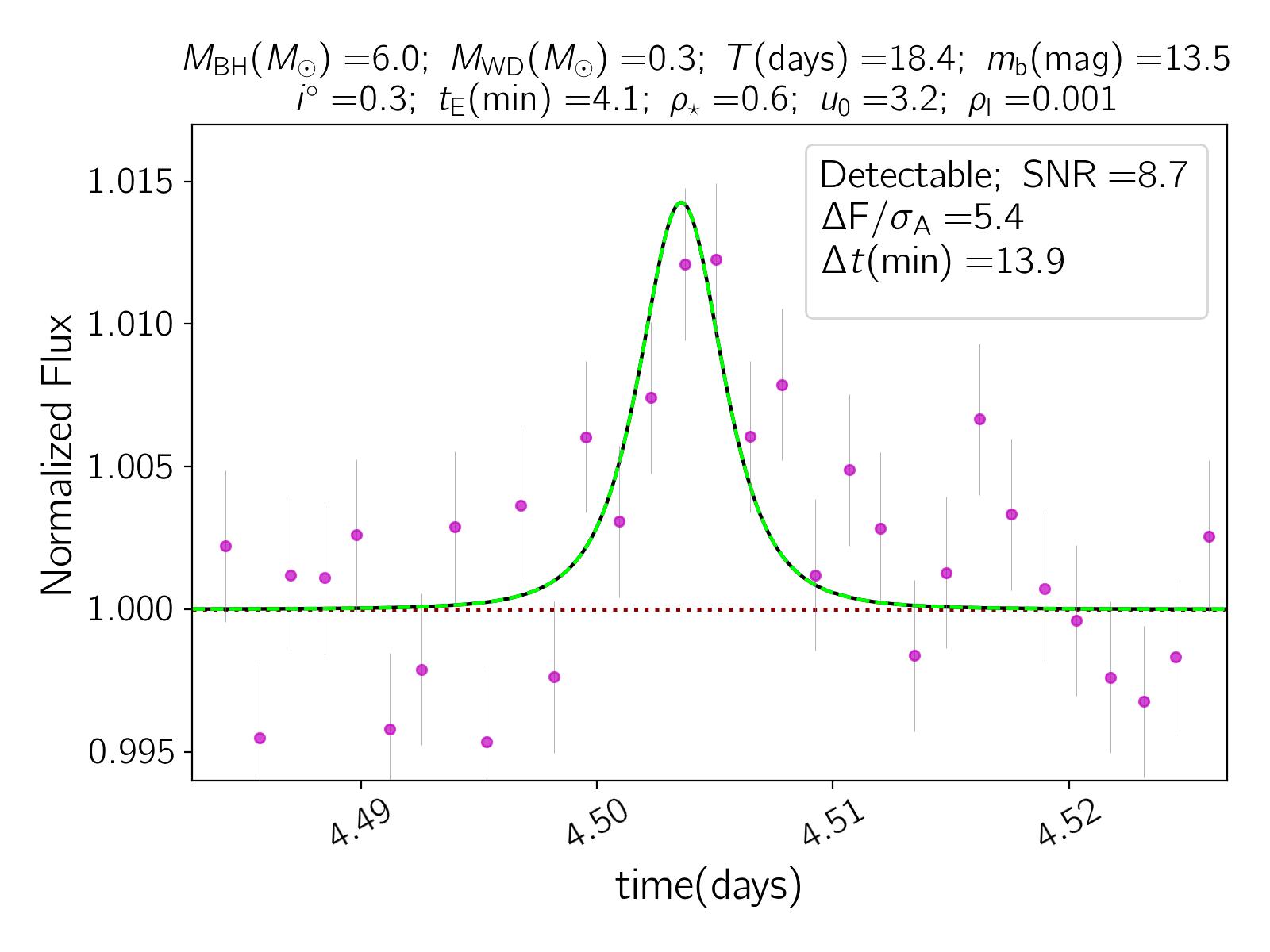}
    \includegraphics[width=0.49\textwidth]{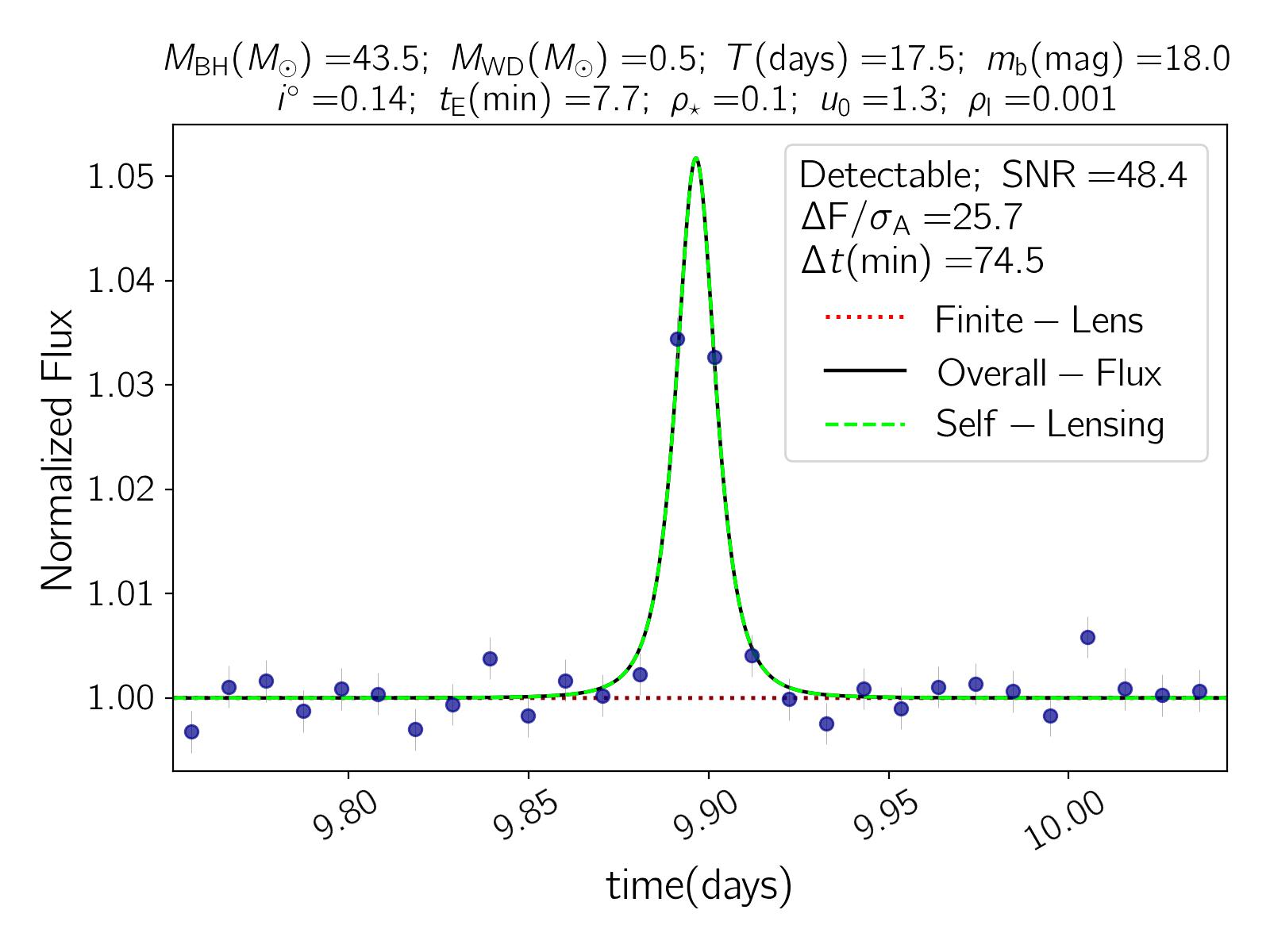}
    \caption{Same as light curves shown in Figure \ref{fig2}, but they are related to three WD+BH binary systems.}
    \label{fig4}
\end{figure}

In Figure \ref{fig4}, we show three examples of simulated light curves due to WD+BH binary systems. The details of these light curves are similar to those shown in Figure \ref{fig2}. For WD+BH binary systems, only one self-lensing signal potentially happens during one orbital period because BHs are dark ($\mathcal{F}=0$) and no self-lensing/eclipsing signals occur for them. Also, there is no occultation for the magnified light of WDs owing to smallness of BHs. The self-lensing signal in the first light curve is not detectable because of its low SNR value. For this event, the depth of signal is in the same order of magnitude as the TESS photometric error. Two next light curves represent detectable self-lensing signals respectively in the TESS and Roman observations. These two light curves have inclination angles ($i=0.1,~0.3^{\circ}$), which is smaller than the inclination angle of the first light curve ($i=0.5^{\circ}$).

\begin{figure}
    \centering
    \includegraphics[width=0.49\textwidth]{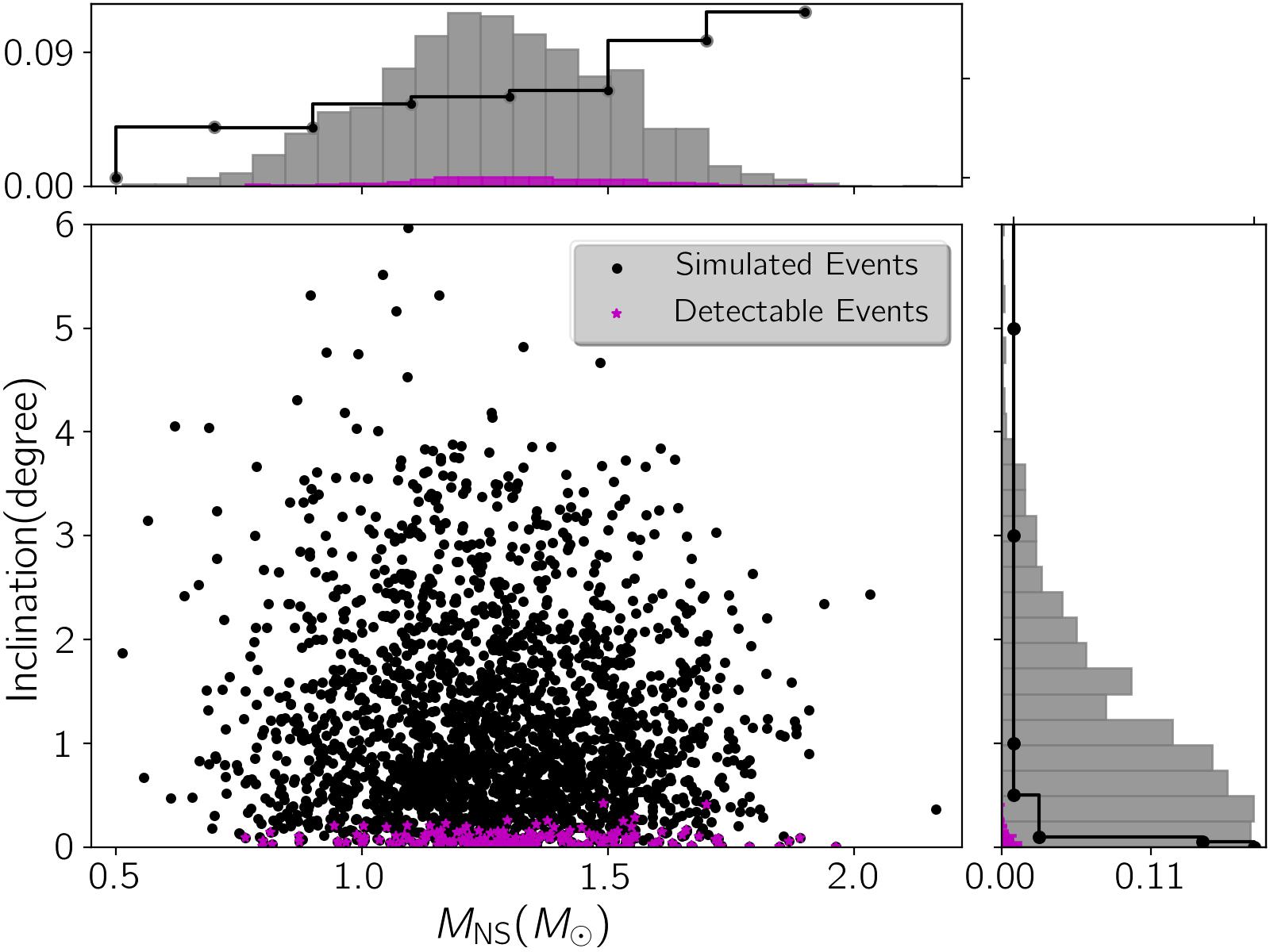}
    \includegraphics[width=0.49\textwidth]{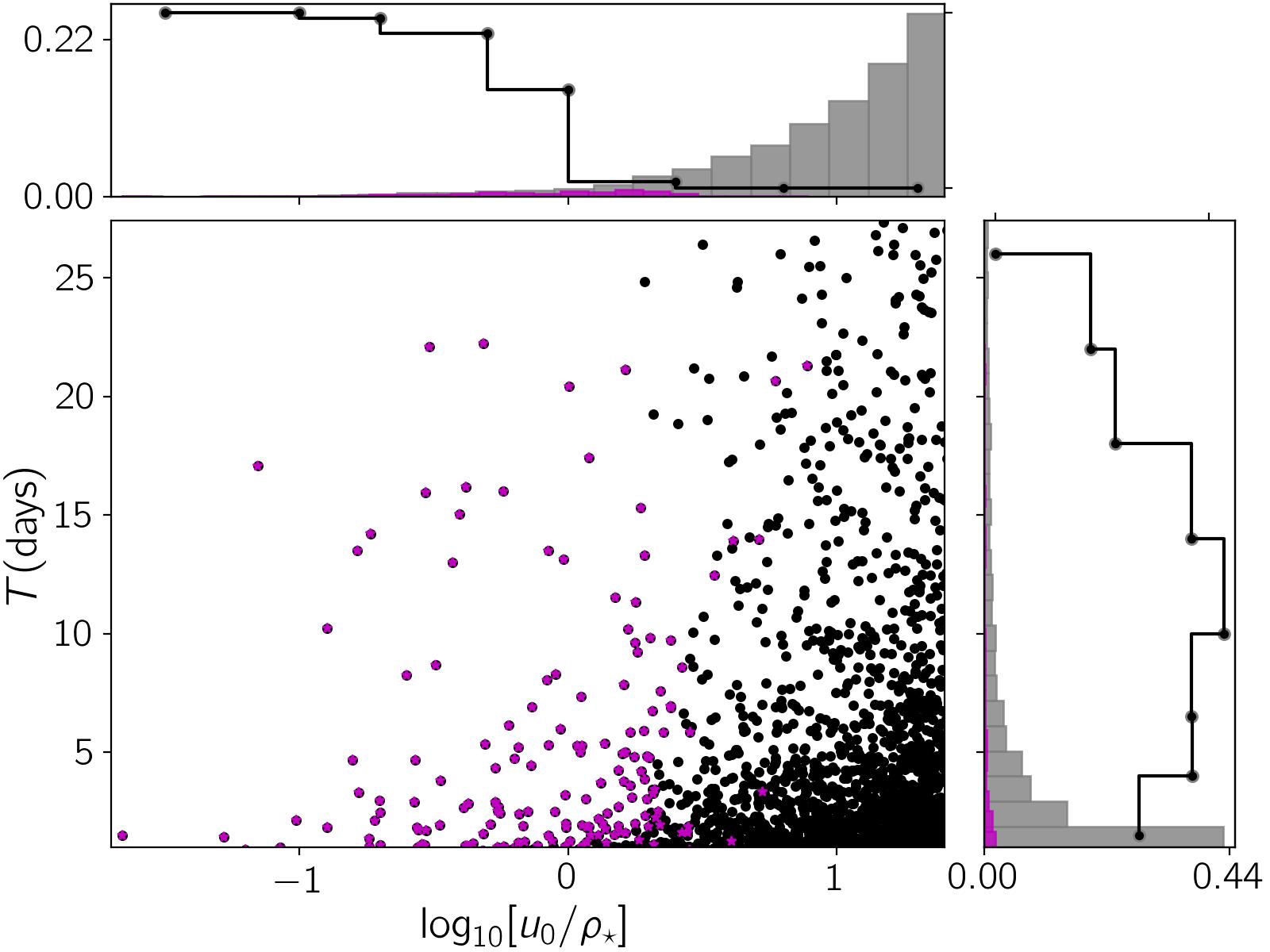}
    \includegraphics[width=0.49\textwidth]{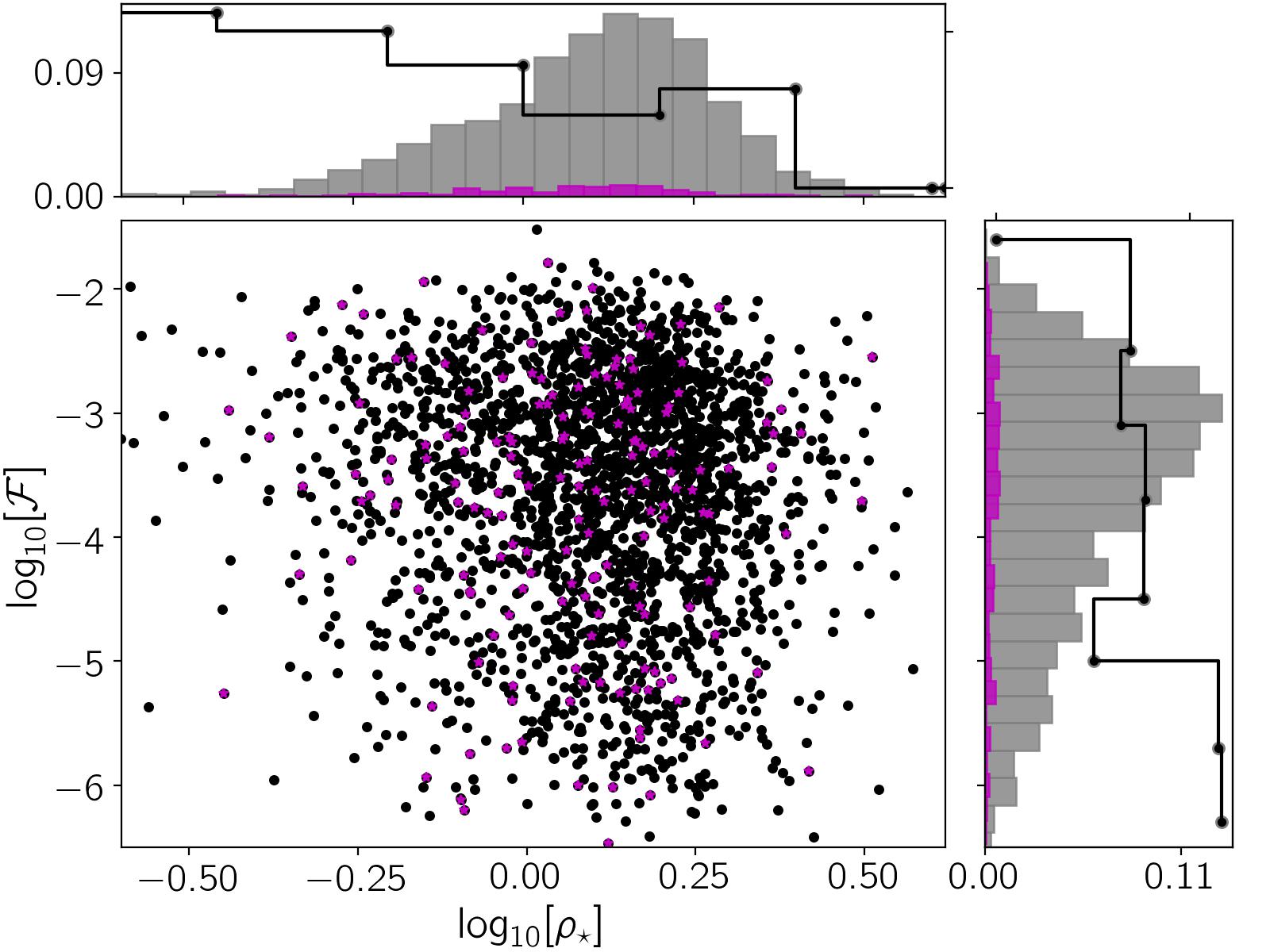}
    \caption{In each map, black dots represent simulated light curves due to WD+NS binary systems over the specified 2D space, and magenta stars display events with detectable signals in the TESS observations. In right and top sides of each panel, the marginal 1D normalized distributions related to the specified parameters over vertical and horizontal axes due to all simulated (and detectable) events are displayed with gray (and magenta) histograms. The solid step lines inside them are the TESS detection efficiencies. Similar plots related to other simulations of binary systems while applying observing strategies can be found in this \href{https://iutbox.iut.ac.ir/index.php/s/kWGPpdB9JYwYpsZ}{link}.}
    \label{fig3}
 \end{figure}

By applying the observing strategies of the TESS and Roman telescopes, we accomplish Monte Carlo simulations for these binary systems. We then evaluate detectability of simulated self-lensing/eclipsing signals, and calculate their statistical parameters, as explained in the following subsection.  

\subsection{Statistical Results}\label{sub33}
Because stars fainter than $17.5~(26)$ mag in the TESS $T$-band (the Roman observations) are not detectable, we exclude them and do not make their light curves. The numbers of these detectable WDs were estimated in Subsection \ref{sub23}. We first select binary systems that in their light curves at least one self-lensing/eclipsing signal occurs during the observing time span (i.e., $1\leq T\leq T_{\rm{obs}}$). Here, we set the low limit (one day) on the orbital period because for binary systems with very short orbital periods the probability of detecting their self-lensing/eclipsing signals is very low. In the RGES observations (which is toward the Galactic bulge) there is another obstacle for discovering these signals which is the blending effect. Hence, in simulations of the Roman observations additionally (i) we select binary systems with $m_{\rm{b}}\in[14.8,~26]$ mag, where $m_{\rm{b}}$ is the baseline magnitude by applying the blending effect, and (ii) consider the blending factor as a probability function. Our method to calculate blending factors can be found in \citet{2019ApJsajadianpoleski,2021MNRASsajadian}. The fraction of all simulated binary systems that have $1\leq T\leq T_{\rm{obs}}$, and additionally for the Roman observations pass the two mentioned criteria (i and ii) is specified by $\varepsilon_{1}$. We label these events as \textit{suitable}, i.e., that  likely have detectable self-lensing signals, and so we generate their light curves and synthetic data points.

Then, we apply three detectability criteria on the simulated light curves to extract the events with detectable periodic (self-lensing or eclipsing) signals as follows.
\begin{enumerate}[leftmargin=3.5mm,label=(\Roman*)]
\item $\Delta F/\sigma_{\rm A}\ge2$, i.e., the signal's depth should be larger than twice the photometric error.
\item $\Delta t\ge\tau$, i.e., the signal's duration should be longer than the applied observing cadence.
\item $\rm{SNR}\ge\rm{SNR}_{\rm{th}}$, i.e., SNR of the self-lensing/eclipsing signal should be larger than the given threshold $\rm{SNR}_{\rm{th}}$.
\end{enumerate}
\noindent We evaluate the SNR value by Equation \ref{snr}, which was first adopted for discerning exo-planetary transiting signals \citep[see, e.g., ][]{2018AJkunimoto}.

In Figure \ref{fig3}, we show the scatter plots of simulated WD+NS binary systems with black dots over three 2D spaces $M_{\rm{NS}}(M_{\odot})-i$, $\log_{10}[u_{0}/\rho_{\star}]-T(\rm{days})$, and $\log_{10}[\rho_{\star}]-\log_{10}[\mathcal{F}]$, from top to bottom. In these panels, the events with detectable self-lensing/eclipsing signals during the TESS observations are specified with magenta stars. For each panel, 1D marginal histograms due to all simulated events (and those with detectable signals) are shown by gray (and magenta) colors, respectively. Inside these marginal histograms, the black step curves represent the TESS detection efficiencies. Similar plots related to other Monte-Carlo simulations from WD+BH binary systems and by applying the Roman observations have the same behaviors.

\begin{deluxetable*}{c c c c c c c c c c c c}
    \tablecolumns{12}
    \centering
    \tablewidth{0.95\textwidth}\tabletypesize\footnotesize
    \tablecaption{Results of simulating light curves related to WD+NS and WD+BH binary systems with detectable self-lensing/eclipsing signals in the specified surveys.\label{tab1}}
    \tablehead{\colhead{System}&\colhead{$\rm{SNR}_{\rm{th}}$}&\colhead{$T$}&\colhead{$i$}&\colhead{$M_{\rm{l}}$}&\colhead{$u_{0}$}&\colhead{$\rho_{\star}$}&\colhead{$D_{\rm l}$}&\colhead{$\Delta t$}&\colhead{$\varepsilon_{1}$}&\colhead{$\varepsilon_{\rm L}$}&\colhead{$\varepsilon_{\rm{tot}}$}\\
        & &$(\rm{days})$&$(\rm{deg})$&$(M_{\odot})$&$(\rho_{\star})$&&$(\rm{kpc})$& $(\rm{min})$&$[\%]$&$[\%]$&}
    \startdata
    \multicolumn{12}{c}{TESS Survey Observations}\\
    WD+NS &$3.0$ & $6.0\pm6.0$ & $0.07\pm0.05$ & $1.3\pm0.2$ & $0.9\pm0.7$ &$1.4\pm0.5$ & $0.13\pm0.07$ &$6.2\pm5.5$ & $17.05$ &$0.14$ &  $2.34e-04$\\
    WD+NS &$6.0$ & $5.7\pm5.9$ & $0.07\pm0.05$ & $1.3\pm0.2$ & $0.9\pm0.6$ &$1.4\pm0.5$ & $0.13\pm0.07$ &$6.1\pm5.6$ & $17.05$ &$0.13$ &  $2.20e-04$\\
    WD+BH &$3.0$ & $6.3\pm6.3$ & $0.18\pm0.12$ & $28.6\pm13.4$ & $7.1\pm6.2$ & $0.2\pm0.1$ & $0.13\pm0.07$ &$16.6\pm13.3$ & $17.39$ & $0.34$ & $5.95e-04$\\
    WD+BH &$6.0$ & $5.9\pm6.0$ & $0.17\pm0.12$ & $28.2\pm13.3$ & $6.3\pm5.1$ & $0.2\pm0.1$ & $0.13\pm0.07$ &$16.0\pm12.8$ & $17.39$ & $0.32$ & $5.57e-04$\\
    \hline
    \multicolumn{12}{c}{Roman Survey Observations}\\
    WD+NS&$3.0$ & $19.2\pm7.3$ & $0.05\pm0.024$ & $1.3\pm0.2$ & $3.5\pm2.1$ & $0.5\pm0.1$ & $0.10\pm0.03$ & $25.1\pm7.1$ & $0.001$ & $0.002$ & $2.69e-10$\\
    WD+NS&$6.0$ & $18.8\pm8.2$ & $0.04\pm0.026$ & $1.4\pm0.2$ & $2.9\pm1.9$ & $0.5\pm0.1$ & $0.11\pm0.03$ & $24.2\pm7.7$ & $0.001$ & $0.001$ & $2.16e-10$\\
    WD+BH&$3.0$ & $18.4\pm16.8$ & $0.10\pm0.09$ & $29.9\pm14.0$ & $7.9\pm7.8$ & $0.1\pm0.0$ & $0.16\pm0.08$ &$33.6\pm21.9$ & $0.002$ & $0.007$ & $1.27e-09$\\
    WD+BH&$6.0$ & $19.9\pm17.4$ & $0.07\pm0.04$ & $31.8\pm14.0$ & $5.7\pm5.9$ & $0.1\pm0.0$ & $0.16\pm0.09$ &$34.4\pm23.3$ & $0.002$ & $0.006$ & $1.11e-09$\\
    \enddata
    \tablecomments{From the third to ninth column, each entry reports the average and standard deviation values of the specified parameter (its column) based on the sample of the simulated binary systems with detectable self-lensing/eclipsing signals. $\varepsilon_{1}$ is the fraction of simulated events that have $1\leq T\leq T_{\rm{obs}}$ and pass the mentioned criteria (i, and ii) while applying the Roman observations, which are suitable and likely have detectable self-lensing signals. $\varepsilon_{\rm L}$ reports the fraction of these suitable events that have detectable self-lensing signals according to three mentioned criteria (I, II, III). The last column reports $\varepsilon_{\rm{tot}}=\varepsilon_{1}\times\varepsilon_{\rm L}$.}
\end{deluxetable*}

In Table \ref{tab1} we report the results of these simulations by considering $\rm{SNR}_{\rm{th}}=3,~6$ that is specified in the second column. From the third to ninth column of this table, the average and standard deviation values of $T$, $i$, $M_{\rm l}$, $u_{0}$, $\rho_{\star}$, $D_{\rm l}$, and $\Delta t$ for simulated binary systems (specified in the first column) with detectable self-lensing/eclipsing signals in the given observing survey are reported. Based on Figure \ref{fig3} and Table \ref{tab1}, we list some key points in the following.

\begin{itemize}[leftmargin=2.0mm]
\item All detectable signals were self-lensing (enhancement in the normalized flux) and the eclipsing effects were not detectable. When the WD is passing behind the NS, the occultation effect is negligible because of smallness of the lens object. When the NS is crossing behind its WD companion, an eclipsing occurs but it makes a very small deviation in the overall flux because $\mathcal{F}\ll1$ (see, e.g., the fourth panel of Figure \ref{fig2}). Therefore, the probability of detecting eclipsing signals in WD+NS and WD+BH binary systems is virtually zero.

\item According to the self-lensing formalism, massive lens objects make higher magnification factors. For that reason the detection efficiency is an increasing function of lens mass. 

\item Detection of self-lensing signals is possible in tightly edge-on WD+NS or WD+BH systems. Therefore, detection efficiency strongly depends on the lens impact parameter and the inclination angle. Light curves with detectable self-lensing signals in the TESS and Roman observations have the average inclination angles $i\lesssim0.2^{\circ}$, as mentioned in the fourth column of Table \ref{tab1}. 

\item We note that magnification factor depends on (i) the lens impact parameter which for a circular orbit is given by $u_{0}=a \sin(i)/R_{\rm{E}}$, and (ii) the finite-source size $\rho_{\star}$. According to the middle panel of Figure \ref{fig3}, events with $\log_{10}[u_{0}/\rho_{\star}]\lesssim1$ potentially have detectable self-lensing signals. Also as mentioned in the sixth column of Table \ref{tab1}, the average values of $u_{0}/\rho_{\star}$ for WD+NS and WD+BH light curves with detectable self-lensing signals are $\sim 1-4,~6-8$, respectively. In our simulations, WD+NS binary systems with impact parameters higher than $\sim8~\rho_{\star}$ were not detectable at all.

\item The orbital period $T$ has two effects on detectability of self-lensing signals, which are (i) for longer orbital periods, the number of transit signals ($N_{\rm{tran}}$) is lower, and (ii) for longer orbital periods the duration of self-lensing signals ($\Delta t$) will be longer. The light curves with detectable self-lensing signals in the TESS and Roman observations have the average orbital periods $\sim 6,~19$ days, respectively.

\item More massive lens objects yield larger Einstein radii, smaller $\rho_{\star}$ values, and as a result higher magnification factors. Therefore, the detection efficiency is higher for events with smaller $\rho_{\star}$ values as shown in the last panel of Figure \ref{fig3}. Also, more massive lens objects (NSs) have less $\mathcal{F}$. Therefore, the detection efficiency is a decreasing function versus $\mathcal{F}$. We note that the applied mass function for NSs had the mean value $1.28M_{\sun}$, while the average value of NSs' mass for light curves with detectable self-lensing signals is $\gtrsim1.3M_{\odot}$ higher than that mean value, as mentioned in Table \ref{tab1}.  

\item Detectable self-lensing signals in WD+BH binary systems do not need very tight alignments between the two components (as seen by the observer) in comparison with detectable self-lensing in WD+NS binary systems (see the fourth column of Table \ref{tab1}). According to our simulation, in the TESS observations WD+BH binary systems with $i\lesssim0.9$ degree can be detected. 

\item WD+BH binary systems with more massive BHs have self-lensing signals with higher detection rate. 

\item The detectable signals in WD+BH binary systems in the TESS and Roman observations last $\sim16,~34$ minutes on average which are longer than detectable self-lensing signals related to WD+NS binary systems. The reason is that WD+BH systems have smaller source radii relative their Einstein radii (see the seventh column of Table \ref{tab1}).   

\end{itemize}

\noindent In Table \ref{tab1}, $\varepsilon_{1}$ represents the fraction of suitable events from all simulated events. We label the events that have $1\leq T\leq T_{\rm{obs}}$, and additionally in the Roman observations their source stars are realizable despite the high blending effects toward the Galactic bulge i.e., they pass the two mentioned criteria (i) and (ii) as the suitable events. Accordingly, in the RGES observations realizing WDs is barely possible. Its main reason is the high blending effect in the RGES observations specially for faint source stars.

\noindent Also, $\varepsilon_{\rm L}$ reports the fraction of the suitable events that have detectable self-lensing signals according to the mentioned criteria (I, II, III). This efficiency for WD+BH binary systems is higher than that for WD+NS binary systems. The shorter cadence of TESS makes higher $\varepsilon_{\rm L}$ in comparison with the similar efficiency for the Roman observations.


\noindent For all simulated WD+NS and WD+BH binary systems the overall probabilities of detecting their self-lensing signals in the TESS observations can be derived by $\varepsilon_{\rm{tot}}=\varepsilon_{1}\times \varepsilon_{\rm L}\sim2-6\times10^{-4}$ which are mentioned in the last column of Table \ref{tab1}. The corresponding probability related to the potential Roman observations during RGES is $\varepsilon_{\rm{tot}}\sim2-12\times10^{-10}$. The main reasons for the low Roman detection efficiency $\varepsilon_{\rm{tot}}$ are (a) its sparse time sampling relative to the durations of these self-lensing/eclipsing signals, and (b) its very crowded field during the RGES observations.

\section{Number of  Detectable Self-Lensing Signals}\label{snew}
Based on the results from the previous sections, we here estimate the number of self-lensing signals due to WD+NS and WD+BH binary systems that can be detected by the TESS and Roman telescopes. 

We derived the number of detectable WDs in the TESS and Roman observations (with $m_{T}\leq17.5$ mag and $m_{\rm{W149}}\leq26$ mag) in Subsection \ref{sub23} and use them here to estimate the number of detectable self-lensing signals in WD+NS and WD+BH binary systems. Accordingly, in the TESS and Roman observations, the numbers of detectable WDs are around $N_{\rm{WD}, T}\simeq55,600$ and $N_{\rm{WD}, W149}\simeq15,000$, respectively. Therefore, the numbers of WD+NS and WD+BH systems with detectable self-lensing signals during the TESS observations from the whole sky (two years) are $\sim N_{\rm{WD}, T}\times f_{\rm b}\times \varepsilon_{\rm{tot}}\sim12-13\times f_{\rm b}, \rm{and}~31-33\times f_{\rm b}$, respectively. Here, $f_{\rm b}$ is the fraction of these WDs that orbit NSs or BHs.

\noindent According to these estimations and if $f_{\rm b}\gtrsim3\%,~\rm{and}~8\%$, at least one self-lensing signal due to WD+BH and WD+NS binary systems are detectable during the TESS observations. We note that the real number of  detectable self-lensing signals by this telescope could be higher/lower somewhat because (i) this telescope observes some parts of the sky over longer durations due to overlapping sectors (self-lensing signals due to binary systems with longer orbital periods can be detected as well), and (ii) its observations are being repeated, and the additional observations can be combined to reduce the overall photometric errors.
 
For the Roman observations, the number of detectable self-lensing/eclipsing signals due to WD+NS or WD+BH binary systems can be estimated as $N_{\rm{WD}, W149}\times f_{\rm b}\times\varepsilon_{\rm{tot}}\sim3-4\times10^{-6}\times f_{\rm{b}},~\rm{and}~2\times10^{-5}\times f_{\rm b}$, respectively. These numbers manifest that detecting self-lensing signals in WD+NS and WD+BH binary systems during one season of RGES is not possible. There are two main reasons for such low Roman efficiency for detecting self-lensing signals due to WD+BH and WD+NS binary systems, which are (i) the Roman cadence is not short enough to capture such short-duration signals, and (ii) the volume of our galaxy that is probed by this telescope during RGES (a cone with the height and radius $5$ and $0.07$ kpc) is too small, and very crowded with high blending effects. The blending effect for observing WDs (that are intrinsically faint source stars) towards the Galactic bulge is a real obstacle.

\section{Conclusions}\label{sec5}
Binary systems from compact objects are potential sources of Gamma-ray bursts, gravitational waves, and super novae explosions. For that reason, they are so-called multi-messenger targets. Therefore, it is valuable to study and detect them. The compact objects in these systems could be WDs, NSs, and BHs. Two classes of these binary systems WD+NS and WD+BH potentially have photometric signals in the visible and infrared passbands in addition to radio pulses or $X$-ray radiations. Through dense photometric observations from these systems one can extract periodic footprints due to their orbital motions. These periodic variations could be due to Doppler boosting, ellipsoidal variation, self-lensing, eclipsing, etc. In this work, we studied the detection and characterization of self-lensing/eclipsing signals in WD+NS and WD+BH binary systems in survey observations that are being done and will be done by the TESS and Roman telescopes.     

For common WD+NS systems, lensing of the WDs by their NS companions generates Einstein radii $R_{\rm E}\sim 0.01R_{\odot}$ similar to a common WD radius, which implies $\rho_{\star}\sim1$. However, $R_{\rm E}$ increases for binary systems with longer orbital periods as $R_{\rm{E}}\propto T^{1/3}$, so that most of WD+NS systems with $T\gtrsim25$ days have $\rho_{\star}\lesssim1$. Very wide WD+NS binary systems with $T\gtrsim150$ days have $\rho_{\star}\lesssim 0.5$ with considerable magnification factors. For WD+BH systems (with stellar-mass BHs $M_{\rm{BH}}\in [3.3,~50]M_{\odot}$) the Einstein radius is larger and reaches to $\sim R_{\odot}$, which implies $\rho_{\star}\in[0.01,~1]$. WD+BH systems with $T\gtrsim3,~20$ days have $\rho_{\star}\lesssim1,~0.5$, respectively.  

We analytically estimated the probability that self-lensing signals in these binary systems offer SNR$\ge6$ in the magnification peak. This probability maximizes for WD+NS and WD+BH binary systems with low-mass WDs revolving massive NS/BH objects. This probability is higher for WD+BH binary systems in comparison with WD+NS systems by at least one order of magnitude.

We generated ensembles of different WD+NS and WD+BH binary systems based on their known distribution functions. We then made their light curves and assumed these light curves to be detected by the TESS and Roman telescopes during the $T_{\rm{obs}}=27.4,~62$-day time spans, with fixed cadences of $\tau=2,~15$ minutes, respectively. We selected the binary systems with $1\leq T\leq T_{\rm{obs}}$ because they could potentially have detectable at least one self-lensing/eclipsing signal. Additionally, in simulations from the Roman observations we applied two criteria to extract realizable WDs against their high blending effects (which were labeled by (i) and (ii), and described in Subsection \ref{sub33}).


To extract detectable self-lensing signals, we applied three criteria to simulated light curves with synthetic data points as SNR$\ge\rm{SNR}_{\rm{th}}$, $\Delta F\ge2\sigma_{\rm{A}}$, and $\Delta t\ge\tau$. We concluded that the fraction of all simulated (WD+NS and WD+BH) binary systems that had detectable self-lensing signals in the TESS observations was $\varepsilon_{\rm{tot}}\sim2-6\times10^{-4}$. In the Roman observations, this fraction was estimated as $\varepsilon_{\rm{tot}}\sim2-12\times10^{-10}$. These efficiencies for WD+BH binary systems are higher than those for WD+NS binary systems by $\sim3$-$5$ times.

The principal reasons for such low detection efficiencies are as follows. (a) The faintness of WDs is an issue so that two binary components should be tightly edge-on to make detectable self-lensing signals among noisy data points. (b) Duration of these signals is short in comparison with the applied observing cadences $\tau=2$, and $15$ minutes. For instance, in our simulations detectable self-lensing signals in the TESS and Roman observations had the average durations $6-16$,~$24-34$ minutes. Hence, in the Roman observations, the number of data points over even detectable self-lensing signals is $1-2$ on average. (c) In the Roman observations, high blending effects towards the Galactic bulge is an extra obstacle to detect all WDs.
   
We finally concluded that detecting at least one self-lensing signal due to WD+NS and WD+BH binary systems in the TESS observations is possible, provided that $\sim8\%~\rm{and}~3\%$ of the Galactic WDs have NS or BH companions. Detecting such self-lensing signals in the Roman observations is impossible because of its long cadence in comparison with the signals' duration, and its small and so crowded fields.\\

\small{We thank the anonymous referee for his/her careful and useful comments, which improved the quality of our paper.}\\

\small{The developed codes for this research, generated animations, figures, and several examples of generated light curves can be found in a Zenodo repository \cite{sajadian2026zenodoit}, and at the GitHub address \url{https://github.com/SSajadian54/WDNS-WDBH}.}\\

\normalsize

\bibliography{ref}{}
\bibliographystyle{aasjournal}
\end{document}